\documentclass[a4paper,12pt]{article} 
\usepackage[T1]{fontenc}

\usepackage[left=4cm,top=3cm,right=2cm,bottom=3cm]{geometry} 
\usepackage[spanish]{babel}

\usepackage[utf8]{inputenc}
\usepackage{amsmath}  
\usepackage{cite}
\usepackage{mathtools}    
\usepackage{microtype}
\usepackage[inline]{enumitem}  
\usepackage[usenames,dvipsnames,table]{xcolor} 
\usepackage{setspace}
\usepackage{atbegshi}
\AtBeginDocument{\AtBeginShipoutNext{\AtBeginShipoutDiscard}}
\begin{document} 
\pagenumbering{arabic} 
\spanishdecimal{.}
\onehalfspacing
\addtocounter{page}{-1}

\title{\textbf{ESTUDIO DE MATERIALES PARA LA FORMACIÓN DE ALEACIONES DE ALTA ENTROPÍA UTILIZABLES EN EL ALMACENAMIENTO DE HIDRÓGENO}}
\author{}
\date{}

 \begin{center} 

\maketitle{JHONN HARY ROYERO BARRAZA}

\vspace{2.3cm}
Trabajo de grado presentado como requisito parcial para optar al título de Físico

\vspace{1.4cm}

\textbf{Director}

Dr. Jairo Plaza Castillo 

\vspace{1.4cm}

\vfill
\includegraphics[width=3cm, height=2.8cm]{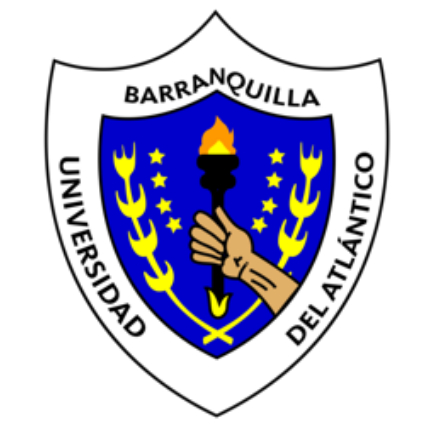} 

UNIVERSIDAD DEL ATLÁNTICO 
 
FACULTAD DE CIENCIAS BÁSICAS 

PROGRAMA DE FÍSICA 

PUERTO COLOMBIA 
  
\textbf{ 2023 }

\end{center}   
\thispagestyle{empty}
\newpage
  \flushleft Nota de aceptación:
  
  El presente trabajo titulado \emph{Estudio de Materiales para la Formación de Aleaciones de Alta Entropía Utilizables en el Almacenamiento de Hidrógeno}, presentado por el estudiante \emph{Jhonn Hary Royero Barraza}, como requisito para optar el titulo de Físico fue aprobado y calificado por los evaluadores como: 
  \rule{60mm}{0.1mm}

  \vspace{4cm}
  Evaluador 1
  
  \vspace{0.4cm}
  \rule{70mm}{0.1mm}
  
\vspace{1.5cm}
  Evaluador 2
  
  \vspace{0.4cm}
  \rule{70mm}{0.1mm}

  \vfill
  \begin{center}
      \textbf{Puerto Colombia, 16 de mayo de 2023}
  \end{center} 

\newpage

\textcolor{White}{y}

\vspace{6cm}
\begin{center}
   Este trabajo de grado se lo dedico a mis padres \emph{José Rafael} y \emph{Ana María}, sin sus esfuerzos e incansables luchas jamás se hubiera podido realizar. 
\end{center}    
\newpage
\tableofcontents 
\newpage
\sloppy 

\setlength{\parindent}{2em}
\setlength{\parskip}{1mm}
\section*{Resumen}
En este trabajo de grado se realizó un estudio teórico computacional sobre las aleaciones de alta entropía (\emph{HEAs} por su traducción del ingles \emph{Hight Entropy Alloys}). Las $HEAs$ son soluciones solidas monocristalinas. Se escogieron dos aleaciones en específico para su estudio, $AlCoCrFeNi$ y $TiVZrNbHf$. El estudio se basó en los parámetros que se deducen de las reglas de Hume-Rothery, la termodinámica y la energía  elástica configuracional (o de mezcla). Se estudió la entalpía de formación como contribución debida a la energía elástica configuracional, con el objetivo de poder conocer cuales son los elementos químicos y sus respectivas  proporciones molares, de tal forma que se favorezca la formación de $HEAs$ y el almacenamiento de hidrógeno. Para lograrlo se hizo un programa en FORTRAN90, esto permitió tener una amplia variedad de resultados que en general están de acuerdo con la literatura teórico-experimental. De manera general se encontró que la aproximación que existe al haber considerado que las \emph{HEAs} son  elásticamente isótropas es correcta, donde se asumió que existe un módulo de compresibilidad (o \emph{Bulk Modulus} por su traducción del ingles) promedio  equivalente para cada especie atómica. También se consideró la variación de la energía elástica configuracional respecto la concentración de la especie atómico de mayor radio atómico,  esto mostró un buen acuerdo con el siguiente criterio que se deduce de las reglas de Hume-Rothery  $\gamma \leq 1.175$, para separar las soluciones solidas  de las múltiples fases cristalinas, fases intermetálicas y vidrios metálicos. $\gamma$ solo depende de los radios atómicos y las concentraciones molares. 

\vspace{0.8cm}
\emph{\textbf{Palabras clave}: Aleaciones de alta entropía HEAs; almacenamiento de hidrógeno; reglas de Hume Rothery; entalpía de formación; energía elástica configuracional.    }

\section*{Abstract}
In this degree work, a theoretical computational study on high entropy alloys (HEAs) was carried out. Two specific alloys were chosen for study, $AlCoCrFeNi$ and $TiVZrNbHf$. The study was based on the parameters deduced from the Hume-Rothery rules, thermodynamics and the configurational (or mixing) elastic energy. The enthalpy of formation was studied as a contribution due to the configurational elastic energy, with the objective of being able to know which are the chemical elements and their respective molar proportions, in such a way that the formation of HEAs and the storage of hydrogen are favoured. To achieve this, a FORTRAN90 program was developed, which allowed a wide variety of results that are generally in agreement with the theoretical-experimental literature. In general, it was found that the approximation that exists by considering that the HEAs are elastically isotropic is correct, where it was assumed that there is an equivalent average Bulk lus for each atomic species. The variation of the configurational elastic energy with respect to the concentration of the atomic species with the largest atomic radius was also considered, this showed good agreement with the following criterion deduced from the Hume-Rothery rules $\gamma \leq 1.175$, to separate solid solutions from multiple crystalline phases, intermetallic phases and metallic glasses. $\gamma$ only depends on the atomic radii and molar concentrations.

\vspace{0.8cm}
\emph{\textbf{Key words}: High entropy alloys HEAs; hydrogen storage; Hume Rothery rules; enthalpy of formation; configurational elastic energy.}
\newpage

\section{Introducción}         
     
 Nuestra sociedad tiene dos vías principales de crecimiento, por un lado se tiene la poblacional (la taza de natalidad debe ser siempre mayor que la taza de mortalidad) y por otro lado se tiene la tecnológica que es necesaria para el mejoramiento de la calidad de vida, conocer y estudiar los fenómenos de la  naturaleza, poder dominarlos y tratar de responder preguntas fundamentales como por ejemplo ¿de dónde venimos? ¿la antimateria rompe la simetría \emph{CPT} en las interacciones débiles?  ¿se puede dar la vida en otro planeta tal y como la conocemos aquí en la Tierra?
       
 Ambas vías implican un aumento de la demanda del  suministro energético que a su vez implica mayor capacidad de almacenar energía, en este sentido las energías alternativas son importantes desde el punto de vista de autosostenibilidad de nuestro planeta, ofreciéndonos una fuente $" inagotable"$ de  energía. Esta autosostenibilidad  involucra dos problemas fundamentales, la producción y almacenamiento de energía; teniendo en cuenta esto, el presente trabajo de grado a tomado la vía del almacenamiento de energía; específicamente se centra en el almacenamiento de hidrógeno por parte de las aleaciones de alta entropía \emph{HEAs} (por su traducción del ingles  \emph{High Entropy Alloys}).    

 Se espera por parte de la sociedad que la transición energética (de combustibles fósiles a energías alternativas) en futuros desarrollos tecnológicos siga favoreciendo el uso del hidrógeno como combustible principal.

Desde aproximadamente el año 2014 se tiene que las teorías para explicar los mecanismos de formación y estabilización de las \emph{HEAs}, se vienen desarrollando  principalmente alrededor de las reglas de Hume-Rothery, la termodinámica  \cite{liang2017} \cite{wang2015} y el exceso de entropía \cite{ding2018}. Para este trabajo de grado se siguió el mismo enfoque, añadiendo, el propuesto por Kamachali y Wang para la energía potencial elástica configuracional \cite{kamachali2022} debido a que el exceso de entropía y la energía potencial elástica pueden ser directamente relacionadas, si consideramos que las \emph{HEAs} son elásticamente isotrópicas. 
      
 De entre todos los modelos que sirven para la predicción de formación de $HEAs$, se destaca el modelo teórico-experimental \emph{CALPHAD} (por sus siglas en ingles  \emph{CALculation of PHAse Diagrams}) para el cálculo de la energía libre de Gibbs, este modelo necesita de constantes  experimentales y a la fecha es el método que más se aproxima a los datos experimentales. La ventaja del enfoque  \emph{CALPHAD} es que a medida que aumentamos el número de componentes de la \emph{HEA}, se necesitan de menos constantes experimentales hasta el punto en que ya no es necesario añadir nuevas constantes, tal como sucede con los sistemas quinarios o superiores, donde las constantes experimentales que se tenían del sistema cuaternario son suficientes. Por otro lado la desventaja del modelo \emph{CALPHAD} radica en los cambios de estructura cristalina, donde la descripción de la energía libre de Gibbs no es la correcta, solo se puede predecir la energía libre de Gibbs con buena precisión mientras no exista un cambio de estructura (por ejemplo de \emph{BBC} a \emph{HCP}) de lo contrario es necesario un nuevo conjunto de constantes experimentales \cite{cacciamani2016}.

 Este trabajo fue enfocado en la búsqueda de \emph{HEAs} con estructura cristalina \emph{BCC} dado que las aleaciones en esta fase son las que tienen mayor capacidad de almacenar hidrógeno, por ejemplo la \emph{HEA} \emph{TiVZrNbHf} con capacidad de hasta $2.5$ $H/M$ ($2.7$ \emph{wt}\% H)\cite{hirscher2020}.
Para entender la tensión de la red y la estabilización en las \emph{HEAs}, tal como lo muestran Kamachali y Wang \cite{kamachali2022}  quienes reajustaron el modelo de Eshelby (originalmente para aleaciones binarias) buscando poder describir la energía potencial elástica de mezcla para aleaciones de múltiples componentes. Este trabajo adoptó el modelo reajustado por Kamachali y Wang  \cite{kamachali2022} para aleaciones multicomponentes, de esta manera se buscó superar la  desventaja presentada por el método   \emph{CALPHAD} cuando existe un cambio de estructura cristalina en la \emph{HEA}.

La tarea 32 de la \emph{International Energy Agency} sobre el almacenamiento del hidrógeno, fue la mayor colaboración internacional en este campo programada para ser desarrollada en el periodo 2012-2019. En ella participaron más de 50 expertos procedentes de aproximadamente 17 países.    
La tarea fue dirigida hacia los materiales porosos, aleaciones intermetálicas e hidruros a base de magnesio como materiales de almacenamiento de energía, hidruros complejos y líquidos de almacenamiento electroquímico de energía, almacenamiento térmico y sistemas de almacenamiento de hidrógeno para aplicaciones fijas y móviles.  Estas investigaciones colectivas no sólo  han dado lugar a más de 600 publicaciones en revistas internacionales y a presentaciones en conferencias y simposios internacionales en este campo, sino que también han permitido descubrir nuevos materiales funcionales \cite{hirscher2020}.      
 
Mediante el desarrollo del presente trabajo se pretendió dar respuesta al
siguiente interrogante:

¿Cuáles son los elementos químicos y sus correspondientes fracciones molares tal que favorezcan la formación de $HEAs$ y la captura de hidrógeno?  

\vspace{0.5cm}

Para la realización de este trabajo se tuvo como objetivo general: 

\begin{itemize}
\item Determinar los elementos químicos y sus proporciones molares, mediante el uso de un
software y el cálculo teórico de la entalpía de formación, tal que favorezcan la formación de $HEAs$ y el almacenamiento del hidrógeno. 
\end{itemize}       
    
Teniendo en cuenta los siguientes objetivos específicos: 1. Seleccionar los parámetros físicos necesarios para la formación de \emph{HEAs} y el almacenamiento de hidrógeno, 2. Calcular con la ayuda del modelo que describe la energía elástica de las \emph{HEAs}  estudiado por Kamachali y Wang \cite{kamachali2022} la función de   partición \emph{Z} para una SS (Solución Solida) de múltiples componentes \cite{takeuchi2001quantitative, he2016configurational} que modele el comportamiento de las mismas y sirva para predecir las  propiedades físicas y químicas de ellas, 3. Determinar con la ayuda de un software las \emph{HEAs} y sus proporciones molares, teniendo en cuenta la máxima capacidad de almacenamiento de hidrógeno impuesta en algunos parámetros debidamente seleccionados; por ejemplo $\delta$ $\leq 0.06 $, donde $\delta$ da la desviación estándar del radio atómico medio, siendo este uno de los limites impuestos tanto para formación de \emph{HEAs} como para la máxima capacidad de almacenamiento de hidrógeno \cite{sahlberg2016}. 4 Comparar los resultados obtenidos por el software, con la literatura teórico experimental.     
   
Este trabajo consta de los siguientes capítulos: \emph{ Introducción}, \emph{fundamentación teórica}, aquí se describen los parámetros termodinámicos que intervienen en la formación de \emph{HEAs}, los que se derivan de la regla de Hume-Rothery, necesarios para la descripción mecánica de éstas, finalmente en este capítulo se plantea un modelo para la entalpía de mezcla haciendo uso de la energía potencial elástica configuracional y la corrección química dada por los coeficientes de Miedema. Seguidamente se pasa al capítulo sobre la  \emph{Metodología}, aquí se presenta el uso de la programación en FORTRAN90 donde se programaron los parámetros mencionados, aparte también se programó la variación de la energía potencial elástica configuracional respecto a la concentración molar de la especie atómica de mayor radio atómico. Se tomaron las siguientes propiedades físicas y químicas, radio atómico, electronegatividad de Pauling, temperatura de fundición, concentración de electrones de valencia, densidad numérica, masa atómica \cite{miracle2017}, módulo de compresibilidad \cite{makino2000estimation}, módulo de cizalladura y el cociente de Poisson \cite{samsonov1968mechanical}. Se estudian en especial las aleaciones $AlCoCrFeNi$ y $TiVZrNbHf$, arrojando como resultado los valores de los parámetros  anteriormente indicados en función de las fracciones molares que componen cada \emph{HEA}. En el capítulo de \emph{Análisis y discusión de resultados}, se presentan los resultados obtenidos de la programación en FORTRAN90 y su comparación con la literatura. Por último se dan las
 \emph{Conclusiones} del trabajo.     

        \newpage
      
\section{Aleaciones de alta entropía (HEAs)} 
  \sloppy    
Los primeros estudios sobre \emph{HEAs} fueron hechos por  Cantor y Yeh \cite{sahlberg2016}, quienes definieron una \emph{HEA} como la mezcla de cinco o más elementos en proporciones aproximadamente equimolares , de acuerdo con Cantor y Yeh cada componente de la  \emph{HEA} en particular, debe estar entre 5 \% y 35 \%, la entropía de mezcla (\emph{$S_{mez}$})  debe ser mayor a  $ 12.47$ $J\cdot mol^{-1}\cdot K^{-1}$ y la entalpía de formación de acuerdo con el modelo de Miedema, debe ser $-15 \hspace{1mm} kJ/mol \leq \Delta H_{mez} \leq 5 \hspace{1mm} kJ/mol$ [Yang and Zhang, 2012, Liang and Schmid-Fetzer, 2017]. Entre los elementos que conforman las \emph{HEAs} se encuentran metales de transición, metales refractarios y metales de tierras raras \cite{shang2021}. De acuerdo con nuestra definición de \emph{HEAs}, estas deben ser mínimo binarias y estar en el estado de SSs (Soluciones Solidas) mono cristalinas (ej. FCC, BCC, HCP)
\cite{liang2017},  El método tradicional de fabricación de estos materiales es el de arco fundición  o aleado mecánico \cite{chen2021} (la idea es alcanzar la temperatura de fundición de la \emph{HEA} para obtener una SS aleatoria/homogénea, ver figura 1b). 

Otra característica de las \emph{HEAs} es que, aunque suelen cristalizar en una estructura simple, a menudo \emph{BCC} (por su traducción del ingles \emph{ Body-Centered Cubic} ), \emph{FCC} (por su traducción del ingles \emph{Face-Centered Cubic}) o \emph{HCP} (por su traducción del ingles \emph{Hexagonal Close Packed}) \cite{shang2021}, sucede que en el proceso de captura de hidrógeno la fase puede cambiar, como sucede para la \emph{HEA} \emph{TiVZrNbHf} con capacidad de almacenamiento de 2.7 \emph{wt \%} (lo que significa que por cada 100 kg de \emph{HEA} se almacenan 2.7 \emph{kg} de $H$ ) para esta \emph{HEA} se encuentra que cuando se alcanza el valor máximo de almacenamiento de hidrógeno, la estructura cristalina pasa de ser \emph{BCC}  a \emph{FCC} o \emph{BCT} (por su traducción del ingles \emph{Body-Centered Tetragonal}) \cite{sahlberg2016}.    

\begin{figure}[htb] 
\centering
\includegraphics[width=0.8\textwidth]{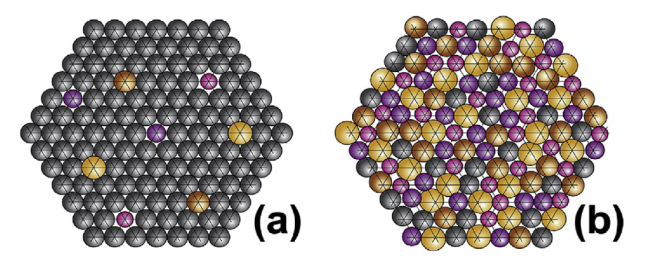}
\end{figure}
 
\begin{center}
\small 
\emph{ \footnotesize \textbf{Figura 1.} El efecto de la diferencia de tamaños atómicos en la posición atómica de las HEAs con 5 especies atómicas diferentes: (a) solución diluida, (b) solución concentrada sin especies de átomos dominantes y posiciones atómicas desviadas de la posición media. La figura (b) es la que más se ajusta a la definición de \emph{HEAs}. Tomado de Miracle y Senkov \cite{miracle2017}. }     
\end{center}   
  
 En el año 2016, Sahlberg  propone que la gran capacidad de almacenamiento de hidrógeno se debe a la tensión de la celda producto de la deformación de la red, que hace que sea favorable la absorción de hidrógeno en los sitios  intersticiales tetraédricos y octaédricos  \cite[página 4]{sahlberg2016}. De acuerdo con Kamachali y Wang, \emph{``... mientras que $\delta$ da la desviación estándar del radio atómico medio $\bar{r}$, los coeficientes de deformación de la red $\lambda_{kh}$ se basan en los desajustes de tamaño atómico entre pares de especies con referencia a un disolvente elegido h."} \cite[página 1]{kamachali2022}.

Los coeficientes $\lambda_{kh}$ son responsables de la tensión en la red, a partir de ellos es posible describir la energía potencial elástica de la \emph{HEA}, esta energía potencial favorece la inestabilidad de la \emph{HEA} (desmezclación) por ende se encuentra compitiendo con la energía entrópica  configuracional o de mezcla $\left ( -T\Delta S^{mez} \right )$ y la entalpía química de mezcla $\left ( \Delta H^{qui} \right )$ cuando ésta es negativa \cite{kamachali2022}.  

A pesar de esto el parámetro $\delta$ toma importancia frente a la energía potencial elástica configuracional  si consideramos que la \emph{HEA} es elásticamente isótropa (esto es, si se considera que la \emph{HEA} tiene un coeficiente promedio de comprensibilidad equivalente para cada una de las componentes atómicas) si esto es así, la energía potencial elástica  configuracional estaría dada como $ e=q\delta^2$, donde $q$ es una constante que depende de los módulos de cizalladuras y los coeficientes de Poisson de las especies atómicas que componen la \emph{HEA}. 
 
La distorsión de la red se calcula con referencia a la constante de red de la solución, normalmente descrita por,

\begin{equation}
\bar{a}=\sum_{k}^{N} c_ka_k=c_1a_1+c_2a_2+...+c_Na_N,
\end{equation} 
siendo $N$ el número de especies atómicas diferentes, $c_k$ y $a_k$ la concentración molar y la constante de red de la $k$ ésima especie atómica. Con base en esto los coeficientes de deformación de la red pueden ser escritos como,  

\begin{equation}
\lambda_{kh}=\frac{1}{\bar{a}} \left ( \frac{\partial \bar{a}}{\partial c_k}  \right )_{c_l}  = \frac{r_k-r_h}{\bar{r}} \approx \frac{a_k-a_h}{\bar{a}}, \hspace{2mm} con  \hspace{2mm} k \neq l \neq h,
\end{equation}  
donde $\sum_{k}^{N} c_k=1 $, $ \bar{a} \approx \bar{r} = \sum_{k}^{N} c_kr_k$. Siendo $r_k$ el radio atómico de la especie atómica $k$, y $h$ el índice de la especie disolvente. En este caso, la elección del elemento disolvente $h$ es arbitraria y no influye en las consideraciones elásticas \cite[página 2]{kamachali2022}. 
  
En el estudio del grado de distorsión de la red de las soluciones solidas, el parámetro,

\begin{equation}
 \delta \%=\sqrt{\sum_{k=1}^{N} c_k \left (1 - \frac{r_k}{\bar{r}} \right )^2}*100\% ,
\end{equation}
  
ha sido popularmente utilizado. El parámetro $\delta \%$ (también conocido como índice de polidispersidad \cite{wang2015} ) se adopta a partir de los debates sobre la cristalización, donde el límite  $\delta \% < 6.66 \%$, se sugiere como criterio para la formación de un cristal estable a partir de un estado líquido. En el caso de las \emph{HEAs}, el parámetro $\delta \% $ ha tenido éxito a la hora de distinguir las soluciones sólidas de las aleaciones amorfas, pero es limitado a la hora de discernir las soluciones sólidas de las aleaciones que forman compuestos intermetálicos o vidrios metálicos, se sugiere que las soluciones solidas y los intermetálicos coexisten en el rango  $4 \% < \delta \%  < 8 \%$ \cite[figúra 2]{wang2015} \cite[figúra 2]{kamachali2022}.

\subsection{Almacenamiento de hidrógeno} 
 
Tradicionalmente, existen tres métodos diferentes de almacenamiento de hidrógeno: 1. gas comprimido, 2. líquido ($H_2$ líquido criogénico o portadores de hidrógeno orgánico líquido), 3. almacenamiento en estado sólido como hidruros  metálicos (véase la Figura 2) \cite{lys2020}.
 
Hasta el día de hoy el almacenamiento gaseoso del hidrógeno es el método más utilizado debido a su relativa sencillez, sin embargo, su baja densidad energética volumétrica a temperatura ambiente y presión atmosférica ($1 kg$ de $H_2$ ocupa $11\hspace{0.6 mm}  m^3$ ) sigue siendo una importante limitación técnica para el uso generalizado del hidrógeno gaseoso. De hecho, se necesita un alto nivel de presurización para cumplir con el requisito de eficiencia volumétrica en los sistemas de almacenamiento de energía a escala industrial, lo que provoca un consumo de energía adicional y  costes adicionales. 

 \newpage

\begin{figure}[!htb]   
\centering
\includegraphics[width=0.98\textwidth]{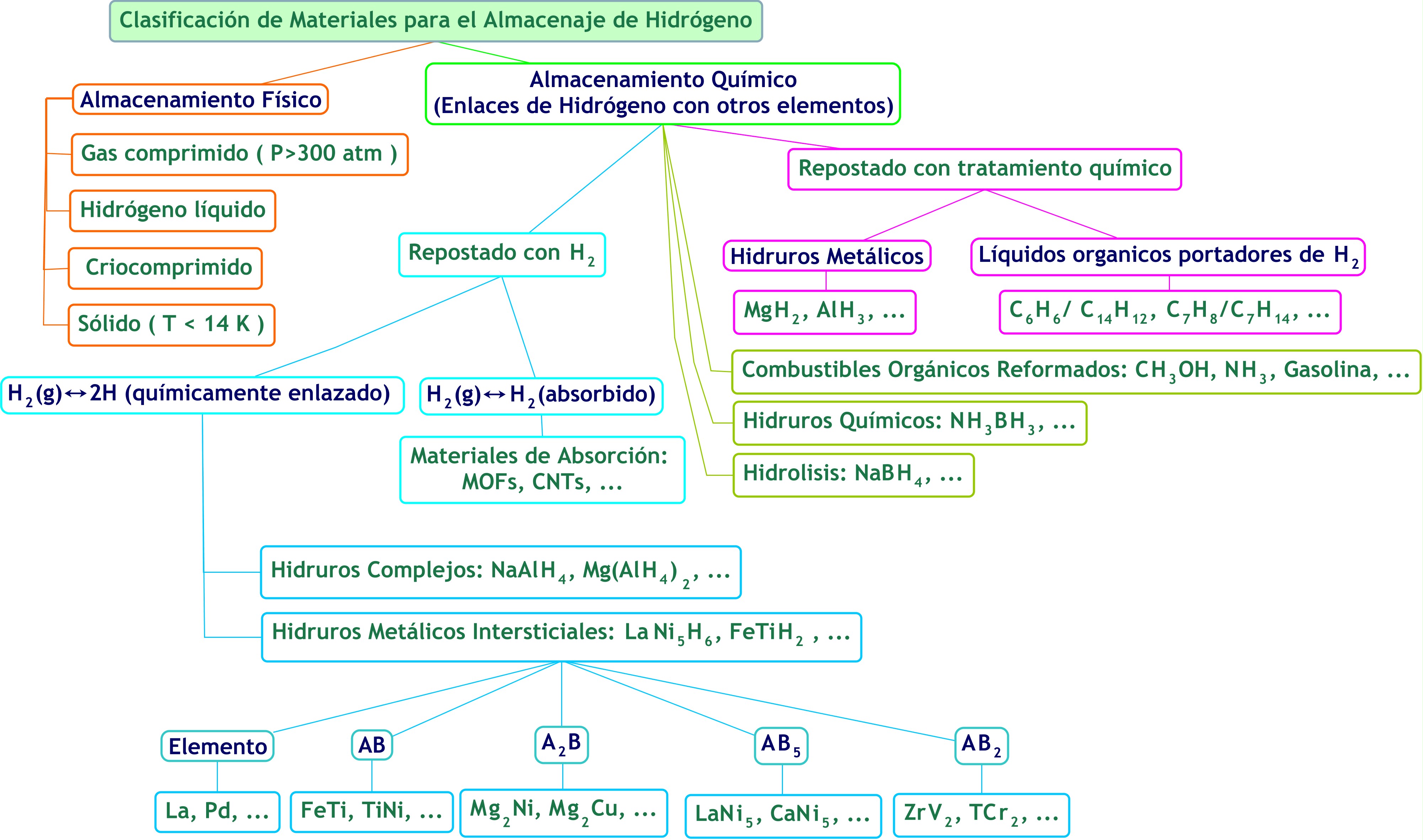} 
 
\end{figure} 

\begin{center} \small 
\emph{ \footnotesize \textbf{Figura 2.} Diagrama de flujo de los materiales y las técnicas para el almacenamiento del hidrógeno. Adaptado de Andrii Lys \cite{lys2020}. }     
\end{center}  
      
El almacenamiento de hidrógeno en estado líquido mejora  las características volumétricas (aproximadamente $40\hspace{0.6 mm} kg$ de $H_2 \hspace{0.6 mm} m^{-3}$ de gas hidrógeno comprimido a $70.8 \hspace{0.6 mm} kg$ de $H_2 \hspace{0.6 mm} m^{-3}$ ), pero requiere condiciones criogénicas ($\sim 21.2 \hspace{0.6 mm} K$ a presión ambiente) para evitar la ebullición (la temperatura crítica del hidrógeno es de $33 \hspace{0.6 mm}K$), o hasta $10^4 \hspace{0.6 mm} atm$ de presión para los sistemas de almacenamiento cerrados a temperatura ambiente. En cualquier caso, el almacenamiento de hidrógeno en estado líquido tiene que superar las barreras técnicas y económicas para las aplicaciones reales, ya que el proceso de licuefacción del hidrógeno (compresión y enfriamiento) consume alrededor del $30\%$ de la energía almacenada y $10^4 \hspace{0.6 mm} atm$ es un reto desde el punto de vista de la ingeniería.

El almacenamiento de hidrógeno en sólidos tiene la ventaja de evitar las mencionadas limitaciones del almacenamiento en estado gaseoso y líquido, y proporciona unas condiciones de almacenamiento seguras y eficaces. El almacenamiento de hidrógeno en estado sólido en hidruros metálicos (HM) parece ser la forma más segura de almacenar hidrógeno, ya que los HM pueden funcionar a temperaturas y presiones relativamente moderadas en comparación con otros estados de almacenamiento. Además, algunos HM pueden considerarse sistemas de almacenamiento térmico, ya que la absorción/desorción de hidrógeno es un proceso exotérmico/endotérmico, respectivamente, que puede desencadenarse fácilmente al operar en condiciones diferentes a las de equilibrio.          
   
Las aleaciones a base de tierras raras también presentan propiedades de almacenamiento (ver figura 2), en el caso más sencillo, los hidruros intermetálicos son compuestos ternarios $A_x B_y H_z$, porque las variaciones en la naturaleza elemental y su cantidad permiten regular las propiedades de sorción y almacenamiento de estos hidruros. El elemento $A$ puede ser una tierra rara o un metal de transición y tiende a formar un hidruro estable. 
El elemento $B$, en cambio, suele ser un metal de transición y no forma hidruros estables. Se ha comprobado que relaciones $B:A$ de 0.5, 1, 2, 5 forman hidruros con una relación hidrógeno-metal de hasta dos. En la tabla (1), se comparan la capacidad de almacenamiento de $H$ y de la densidad de energía almacenada de algunos compuestos y aleaciones \cite{lys2020}.  
    
 \vspace{0.6cm}   
\begin{center}   
\begin{tabular}{|c|c|c|c|c|} \hline  
\multicolumn{1}{|m{3cm}|}{\centering \textbf{Material} } & \multicolumn{1}{m{1.7cm}}{ \textbf{Densidad}} &  \multicolumn{1}{m{1.3cm}}{\raggedright \textbf{de H} } & \multicolumn{1}{|m{1.8cm}}{ \raggedleft \textbf{Densidad} } & \multicolumn{1}{m{2.3cm}|}{ \raggedright \textbf{de energía} }  \tabularnewline     
 & wt\% & $kg \cdot m^{-3}$ & $MJ\cdot kg^{-1}$ & $MJ\cdot dm^{-3}$ \\ \hline
Gas $H_2$, 700 bar & 100 & 42 & 120.0 & 5 \\ $H_2$ liquido (20 K)

 &100 & 71 & 120.0 & 8.5 \\  
$ La Ni_5 H_6$ & 1.4 & 90 & 1.7 & 10.8 \\  
$Ti Fe H_2$  & 1.9 & 105 & 2.3& 12.6 \\   
 $Mg H_2$& 7.6 & 110 & 9.2 & 13.3 \\ 
\hline 
\end{tabular}
\end{center} 
\begin{center} \small
\emph{\footnotesize\textbf{Tabla 1.} Capacidades de almacenamiento de hidrógeno y energía de algunos compuestos. Tomado de Lys \cite{lys2020}. }    
\end{center}   
 
Entre las diferentes clases de aleaciones (soluciones solidas, intermetálicos, vidrios metálicos entre otros), las aleaciones refractarias con red $BCC$ y las sustituciones relacionadas con elementos ligeros pueden llegar  a almacenar  grandes cantidades de hidrógeno $\approx 2.5 \hspace{0.8mm}wt$ $\%$. Se propone la aleación $BCC$ $TiVZrNbHf$ como un material prometedor con un mejor rendimiento de almacenamiento de hidrógeno \cite{hirscher2020}.  

Aparte de que las estructuras BCC tienen una gran capacidad de almacenamiento de hidrógeno \cite{sahlberg2016} también es conocido que la tensión de la red producida por la variedad de radios atómicos diferentes (ecuación 2) favorece el almacenamiento de hidrógeno (formación de HM). Estos dos hechos sugieren que las \emph{HEAs} son excelentes candidatos para el almacenamiento de hidrógeno.
La formación de SSs es favorecida por la alta entropía que a su vez es favorecida por el desajuste de la red que produce la diferencia de tamaños atómicos. Por ejemplo para la aleación $ZrTiVCrFeNi$ \cite{sahlberg2016} con una capacidad del $1.81$ $wt\%$, las propiedades de almacenamiento de hidrógeno se observaron a $100$ $bar$ y $50$ $^oC$ después de una activación de $500$ $^oC$ de síntesis. Después de otros tratamientos térmicos la máxima capacidad de almacenamiento de hidrógeno disminuyo a $1.56$  $wt\%$ bajo las mismas condiciones de hidrógeno. Por otro lado la aleación $TiZrNbMoV$ de acuerdo con Sahlberg \cite{sahlberg2016} presentó una baja capacidad de almacenamiento de hidrógeno ($85$ $bar$ y $50$ $^oC$) cerca de $0.59$  $wt\%$ en las estructuras monocristalinas de tipo BCC como comparación a las estructuras de múltiples estructuras cristalinas ($2.3$ $wt\%$).
             
Para la aleación $TiVZrNbHf$ según Sahlberg  \cite{sahlberg2016} se encontró que este tipo de aleaciones cristalizan en estructuras de tipo BCC acorde con los resultados de este trabajo ($VEC<6.87$), ellos predijeron  un límite máximo del $\delta\%=6.82\%$ (este limite también mide la máxima capacidad de hidrógeno), en nuestro trabajo de grado  aparte de que se reproduce el mismo resultado, se predijo un limite máximo de $\delta\%=8.726\%$, para la aleación $Ti_{0.05}V_{0.35}Zr_{0.35}Nb_{0.05}Hf_{0.20}$, es curioso que se necesite la misma proporción atómica de $V$ y $Zr$, porque el $V$ es el elemento químico de menor radio atómico ($131.6$ $pm$) y el $Zr$ es el de mayor radio atómico ($160.25$ $pm$) \cite{miracle2017}. La absorción del hidrógeno la hicieron de forma isotérmica a $299$ $^oC$, este dato fue grabado incrementando la presión paso a paso alrededor de $50$ $bar$, la absorción de la cinética fue baja y el tiempo de equilibrio fue un conjunto de máximo por punto de la isoterma. Ellos encontraron con ayuda de técnicas de mediciones gravimétricas, que, la aleación $TiVZrNbHf$ absorbía fácilmente el hidrógeno por encima de los $200$ $^oC$. Un tratamiento de $400$ $^oC$ por $48$ $h$ llevó a una hidrogenación completa de la muestra, con un cambio de estructura $BCC\rightarrow BCT$, para un almacenamiento de hidrógeno superior al $H/M>2.3$ \cite{sahlberg2016}.    

Como aplicación a los sistemas de almacenamiento de hidrógeno en las  figuras (3, 4 y 5)  se presentaron la aplicación para un montacargas pequeño o carretilla elevadora, para el almacenamiento se usaron aleaciones a base de tierras raras del tipo $C14-AB_2$ \cite{hirscher2020}, 

\begin{figure}[!htb]   
\centering
\includegraphics[width=0.7\textwidth]{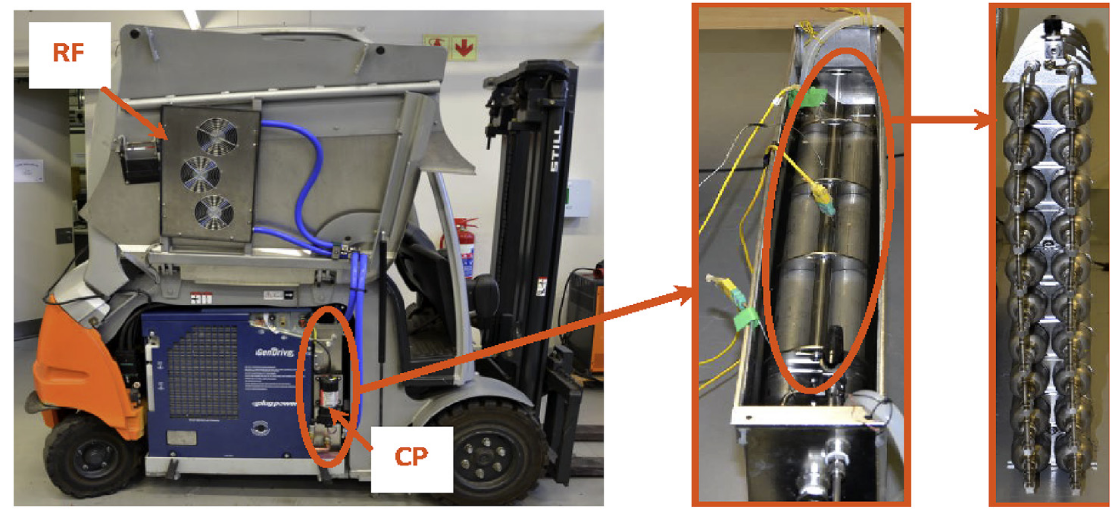} 
 
\begin{center} \small 
\emph{ \footnotesize \textbf{Figura 3.} Izquierda: Montacargas o carretilla elevadora eléctrica equipada con módulo de potencia de pila de combustible y depósito de extensión de almacenamiento de hidrógeno de hidruro metálico (rodeado por un círculo). Centro: el depósito de HM. Derecha: montaje de los contenedores de HM. Tomado de Hirscher \cite{hirscher2020}.}    
\end{center} 
\end{figure}

\begin{figure}[!htb]   
\centering
\includegraphics[width=0.7\textwidth]{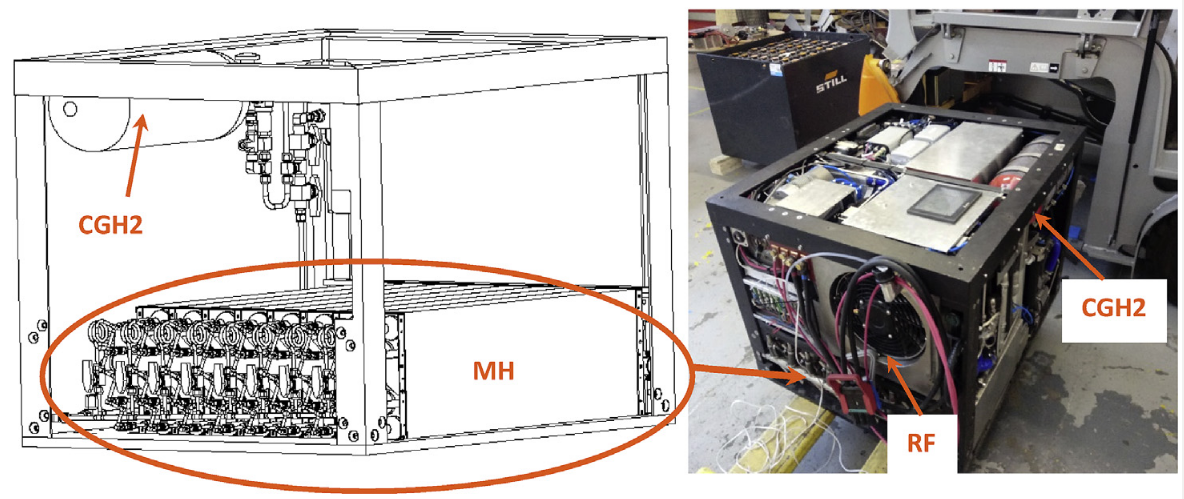} 
 
\begin{center} \small 
\emph{ \footnotesize \textbf{Figura 4.} Depósito de HM (o MH por sus traducción del ingles Metal Hydride) con cilindro amortiguador CGH2 de 9 $L$ integrado en el módulo de potencia HySA Systems para un montacargas eléctrico. Tomado de Hirscher \cite{hirscher2020}.}    
\end{center} 
\end{figure}

\begin{figure}[!htb]   
\centering
\includegraphics[width=0.7\textwidth]{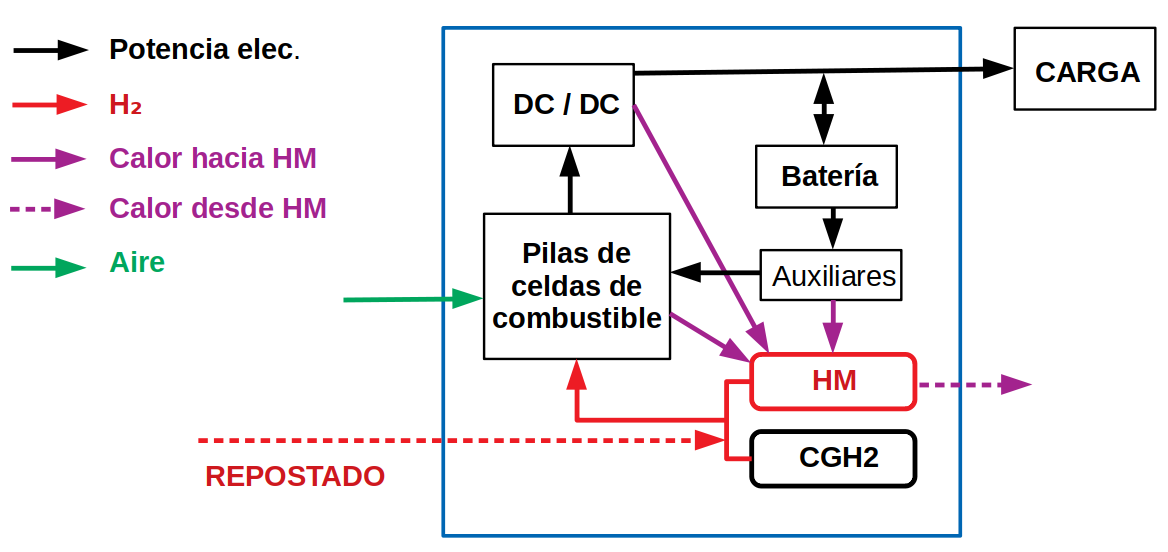} 
 
\begin{center} \small 
\emph{ \footnotesize \textbf{Figura 5.} Vehículo de manipulación de materiales con pila de combustible y almacenamiento de hidrógeno a bordo: un concepto de HySA Systems. Tomado de 
 Hirscher \cite{hirscher2020}.}    
\end{center} 
\end{figure}

Los HM que funcionan a temperatura ambiente o en torno a ella, proporcionan capacidades de almacenamiento de hidrógeno cercanas al 2 $\%$ en peso de H (2 $wt\%$) H (es decir, el almacenamiento de 1 $kg$ de H requiere aproximadamente 50 $kg$ del material HM) pueden proporcionar el almacenamiento compacto de $H_2$ necesario para los vehículos utilitarios pesados impulsados por pilas de combustible (PC) y las carretillas elevadoras o montacargas, en los que el sistema de almacenamiento de hidrógeno puede servir al mismo tiempo como lastre.

Así pues, la limitada capacidad gravimétrica de almacenamiento de hidrógeno de los hidruros metálicos, que a menudo se considera una gran desventaja para su uso en el almacenamiento de hidrógeno vehicular, se convierte en una ventaja en una aplicación de este tipo. 

En las figuras (3, 4 y 5) se muestra la integración con éxito del almacenamiento de hidrógeno de HM de los módulos de alimentación de PC para carretillas elevadoras eléctricas.
HySA Systems, de la Universidad de Western Cape, Sudáfrica, ha demostrado su eficacia. El concepto del sistema (figura 5) se basa en una solución híbrida distribuida de almacenamiento de hidrógeno en la que un tanque de almacenamiento de hidrógeno de hidruro metálico (HM) está conectado a un tanque de hidrógeno de gas comprimido (CGH2 por su traducción del ingles Compressed
Gas Hydrogen) utilizado como amortiguador para las oscilaciones de presión del $H_2$ durante el funcionamiento de la PC, así como para proporcionar una fuerza motriz de presión suficiente para la absorción de $H_2$ en el HM durante los breves periodos de repostaje del sistema. El depósito de HM está acoplado térmicamente con la PC y los componentes del balance de planta, generan calor para equilibrar la desorción endotérmica del HM. En el modo de repostaje, el calor liberado durante la absorción exotérmica de $H_2$ en el MH se disipa en el medio ambiente. 

En la figura (3) se muestra el primer prototipo del sistema de almacenamiento de hidrógeno HM de HySA Systems integrado con un módulo de alimentación PC comercial, instalado en una carretilla elevadora eléctrica estándar de 3 toneladas. El depósito de extensión de almacenamiento de hidrógeno HM se construye como un conjunto de 20 contenedores de HM sumergidos en un depósito de agua. El colector de gas del depósito está conectado al lado de alta presión del cilindro de amortiguamiento (cilindro compuesto de 74 L, CGH2). La gestión del calor del depósito de HM se realiza mediante una mezcla de agua y glicol en circulación accionada por una bomba de circulación (CP \emph{Circulation Pump}) y un conjunto de radiador-ventilador (RF \emph{Radiator Fan}) e intercambiador de calor aire-líquido. El depósito de HM fue optimizado de tal manera que proporcionara una fácil activación y una rápida carga/descarga de $H_2$. El sistema tiene la misma capacidad de almacenamiento de hidrógeno (1.7 $kg$) que el depósito CGH2 cargado a una presión de  350 $bar$, pero a una presión de carga de $H_2$ inferior (185 $bar$).

En la figura (4) se muestra un módulo de potencia de PC de 15 $kW$ desarrollado recientemente por \emph{HySA Systems}, que utiliza un segundo prototipo del depósito de almacenamiento de hidrógeno. Dado que todo el balance de planta se hizo a medida, permitiendo flexibilidad en la integración del depósito de HM en el sistema de PC.
Los tanques de HM se caracterizaron por una capacidad de almacenamiento de hidrógeno de unos 170 $NL/g$, una inclinación en las isotermas de presión-composición (presión de equilibrio del $H_2$ a $RT$ entre 5 y 10 $bar$) y un valor absoluto bajo de entalpía de hidrogenación $( \Delta H= - 18.5$ $kJ/mol$ $H_2$, $\Delta S= - 78.1 J/(mol$ $H_2$ $K))$, lo que minimiza la liberación de calor durante el repostaje y relaja el requisito de suministro de calor durante la liberación de $H_2$ a la pila de PC.

El sistema tiene 40 contenedores de HM (capacidad total de almacenamiento de $H_2$ de unos 1.7 $kg$). Otras características del prototipo inicial son (i) integración directa del sistema de calefacción/refrigeración del depósito de HM con la pila de FC; (ii) contrapeso mediante el encapsulado de los contenedores de HM en plomo fundido y solidificado para alcanzar el peso total requerido, y (iii) un menor tamaño del amortiguador de gas (9 $L$, CGH2 en la figura 4) en comparación con el primer prototipo (74 $L$). Las pruebas de carga pesada realizadas recientemente con la carretilla elevadora suministraron hasta 170 $NL/min$ de $H_2$ a la pila de PC (potencia media de unos 14 kW), a 6-12 $bar$ y calentando a temperaturas de hasta 55 $^oC$. El tiempo de repostaje del depósito de HM a temperaturas ambiente -15 $^oC$ y 20 $^oC$ fue de entre 15 y 20 $min$. 

La principal ventaja de los sistemas de almacenamiento de hidrógeno que utilizan MH reside en una menor presión de almacenamiento del hidrógeno en comparación con la opción de almacenamiento CGH2. Según una estimación reciente, la sustitución de un tanque de almacenamiento CGH2 por uno HM a bordo de un vehículo PC permite reducir aproximadamente un 38 $\%$ los costes de repostaje debido a una reducción significativa de los costes de compresión del hidrógeno \cite{hirscher2020}.

\subsection{Parámetros en la formación de HEAs }   

Las reglas empíricas propuestas en la literatura \cite{liang2017} para predecir la estabilidad de la estructura cristalina de las SSs multicomponentes se dividen esencialmente en dos grupos: el primer grupo se basa en las reglas de Hume-Rothery, tales como los parámetros de diferencia de tamaño atómico $(\delta, \gamma)$, electronegatividad $(\Delta \chi)$ y concentración de electrones de valencia \emph{VEC} (por su traducción del ingles \emph{Valence Electron Concentration}); el segundo grupo utiliza parámetros que se derivan de las propiedades termodinámicas, como, entalpía configuracional o de mezcla $(\Delta H_{mez})$, entropía configuracional $(\Delta S_{mez})$, ideal $(\Delta S_{id})$,  y corregida o correlacionada $(\Delta S_{corr})$ \cite{ding2018}.  Por último se tomó la condición propuesta por Kamachali y Wang \cite{kamachali2022} para las \emph{HEAs} consideradas elásticamente isótropas, donde se consideró la variación de la energía potencial elástica configuracional ($\Delta e$) respecto a la variación de la $i$ ésima concentración atómica ($\Delta c_i$), esto es, $\left.\frac{\Delta e}{ \Delta c_{i}}\right |_j < 0.16$, aquí la barra denota la sustitución $i \rightarrow j$ entre pares de especies atómicas diferentes, explicada más adelante en el capitulo sobre la \emph{metodología}  (capitulo 3).

\subsubsection{Parámetros que se derivan de las reglas de Hume-Rothery}

Las reglas de Hume-Rothery, establecidas en 1938 por el químico-metalúrgico, William Hume-Rothery, representan un conjunto de condiciones que nacen de la observación, estas condiciones deben ser cumplidas por las SSs sustitucionales metálicas binarias para que tenga lugar la miscibilidad total entre los distintos componentes. Dichas reglas establecen
que: 

\begin{itemize}

    \item La diferencia entre los diámetros atómicos debe ser
    15 \%.

    \item La estructura cristalina del solvente y el soluto deben ser iguales o similares. 
 
    \item Valencia (capacidad de ceder electrones a la nube electrónica) con la que actúan ambas especies debe ser igual o similar.

    \item La electronegatividad (capacidad del átomo para atraer electrones) debe ser similar entre ambas especies.
    
\end{itemize}

Con base en las reglas de Hume-Rothery se ha propuesto un conjunto de nuevos parámetros aplicables a las \emph{HEAs}  \cite{liang2017},

\begin{equation}
\Delta \chi =\sqrt{\sum_{i=1}^{N} c_i (\chi_i - \bar{\chi})}, 
\end{equation}

\begin{equation}
 \delta =\sqrt{\sum_{i=1}^{N} c_i \left (1 - \frac{r_i}{\bar{r}} \right )^2} ,
\end{equation}
\begin{equation}
\gamma=\frac{1-\sqrt{\frac{(r_{min} + \bar{r})^{2}-\bar{r}^{2}}{(r_{min} + \bar{r})^{2}}}}{1-\sqrt{\frac{(r_{max} + \bar{r})^{2}-\bar{r}^{2}}{(r_{max} + \bar{r})^{2}}}} ,
\end{equation} 
\begin{equation}
 VEC = \sum_{i=1}^{N} c_i VEC_i ,    
\end{equation}   
  
donde, $ \Delta \chi $ es la diferencia de electronegatividad en el sentido de Pauling, $\bar{\chi}=\sum_{i=1}^{N} c_i \chi_i$, $\chi_i$ es la electronegatividad de la $i$ ésimo componente de la \emph{HEA}, $\delta$ mide el desajuste atómico (ver ecuación 3), $\gamma$ está relacionada a la solubilidad de la \emph{HEA} y es importante porque sirve para diferenciar SSs de vidrios metálicos y aleaciones intermetálicas \cite{wang2015}, $VEC$ es la concentración de electrones de valencia, $VEC_i$ es el $VEC$ de la $i$ ésima especie atómica, $\bar{r}$ es el radio promedio y se define como, $\bar{r}=\sum_{i=i}^{N} c_i r_i$ , aquí $r_i$ es el $i$ ésimo radio atómico, $c_i$ es la $i$ ésima concentración molar atómica, $r_{min}$ es el radio atómico del átomo mas pequeño, $r_{max}$ es el radio atómico del átomo más grande.  Se propone que la gran capacidad de almacenamiento de hidrógeno se debe a la deformación de la red de la aleación, que hace que sea favorable la captura del hidrógeno tanto en sitios intersticiales tetraédricos como octaédricos \cite{sahlberg2016}. Un valor alto de $\delta\%$ conduce a una gran distorsión de la red y hace que la formación de \emph{HEA} sea menos favorable y por ende se pierda la capacidad de almacenamiento de hidrógeno. Se propone que las \emph{HEAs} se forman cuando $\delta\%  \leq 6.6\%$, pero para el $TiVZrNbHf$ (estructura \emph{BCC}) este valor es $\delta\%= 6.8\%$ \cite{sahlberg2016}. Debido a que   las estructuras cristalinas \emph{BCC} son las que mayor capacidad de almacenamiento de hidrógeno poseen \cite[página 12]{hirscher2020}, para este trabajo se tomó como condición para la formación de este tipo de estructuras, $VEC < 6.87$ \cite{liang2017}.

\subsubsection{Parámetros que se deducen de la termodinámica} 

Los parámetros que se deducen de la termodinámica son, la entropía de mezcla $S_{mez}$, la entalpía de mezcla $H_{mez}$, la energía libre de Gibbs de la mezcla $G_{mez}$ y la temperatura de fundición $T_f$ de la \emph{HEA}, y se definen así \cite{liang2017},     

\begin{equation}
 \Delta S_{mez} = -k_B \sum_{i=1}^{N} c_i ln{(c_i)},
\end{equation}
\begin{equation}
 \Delta H_{mez}= 4 \sum_{i}^{N-1}\sum_{ j> i}^{N} \Delta H_{ij}  c_i c_j ,
\end{equation}
\begin{equation}
 T_{f} =  \sum_{i=1}^{N} c_i T_{fi},
\end{equation}
\begin{equation}
  \Delta G_{mez} = \Delta H_{mez} - T_f \Delta S_{mez}.
\end{equation} 

Aquí $k_B$ es la constante de Boltzmann, $\Delta H_{ij} $ es un factor binario que se deduce del modelo de Miedema \cite{takeuchi2005classification}, este mide la diferencia entre entalpías de mezcla entre cada par de átomos,
$T_{fi}$ es la temperatura de fundición de la $i$ ésima especie atómica que conforma la \emph{HEA}.

Para una \emph{HEA} quinaria se tiene que la entropía máxima configuracional ($N_A \Delta S_{mez}$, $N_A:$ número de Avogadro) es $ 1.61 R$ y la mínima es $1.36 R$, algunos autores consideran que los sistemas cuaternarios presentan una entropía ideal ($S_{id}$) igual a $1.39 R$, por otro lado es común encontrar que la definición más adecuada para las \emph{HEAs} en cuanto a la entropía, son aquellas que poseen $S_{id}\geq 1.5 R$ \cite{miracle2017}, de acuerdo con Ding  como el número de componentes $N$ incremente de 3 a 6, así aumenta la entropía ideal $S_{id}$ de $1.1 R$ a $1.79 R$ \cite{ding2018}.  

Para este trabajo se tomó la siguiente combinación de los parámetros ya mencionados, 
\begin{equation}
 \Omega = \frac{T_f  \Delta S_{mez}}{|\Delta H_{mez}|},
\end{equation}    
\begin{equation}
 \Phi = \frac{\Delta S_{mez} - \Delta S_{H}}{|\Delta S_E|},
\end{equation}
\begin{equation}  
 \Delta S_{H}=\frac{|\Delta H_{mez}|}{T_f},
\end{equation}
\begin{equation}
\Lambda = \frac{\Delta S_{mez}}{\delta ^{2}},  
\end{equation}  
siendo $\Delta S_E$, el exceso de entropía, definida más adelante en el capítulo (2.4). En resumen, de la literatura se tomaron para este trabajo las siguientes condiciones impuestas a los parámetros ya mencionados (ver tabla 2),  necesarios para la formación de \emph{HEAs}, 

 \begin{center}
\begin{tabular}{|c|c|c|c|c|} \hline 
\multicolumn{1}{|m{2cm}}{\centering \textbf{Parámetro} } & \multicolumn{1}{|m{2cm}|}{\centering \textbf{Unidad} } &  \multicolumn{1}{m{3.55cm}|}{\centering \textbf{Rango de valores} } & \multicolumn{1}{|m{2.5cm}|}{\centering \textbf{Estructura} } & \multicolumn{1}{m{2cm}|}{\centering \textbf{Fuente} }  \tabularnewline \hline
$\delta \%$ & & $\leq 6.6 \% $ & & $*$ \\
$\gamma$ & & $\leq 1.175 $ & & $**$ \\

$\Delta \chi $ & & $ 0.1\leq \Delta \chi  \leq 0.15$ & & $***$ \\  
$ VEC$ & $e/mol$ & $ \geq 8 $ & FCC& $***$ \\  
$ VEC$ & $e/mol$ & $ < 6.87 $ & BCC& $***$ \\  
$ VEC$ & $e/mol$ & $ 6.87  \leq VEC < 8$  & FCC + BCC& $***$ \\ 
$ \Delta H_{mez}$ & $kJ/mol$ & $-15  \leq \Delta H_{mez} \leq 5  $ & & $***$  \\ 
$ \Delta H_{mez}$ & $kJ/mol$& $-11.6 \leq \Delta H_{mez} \leq 3.2 $ & & $***$\\ 
$ \Omega $ & & $ > 1.1 $ & & $***$ \\ 
$ \Phi $ & & $ > 20 $ & & $***$ \\
$ \Lambda $ & & $ > 0.95 $ & & $***$ \\
\hline
\end{tabular}
\end{center}

 \begin{center}
\small
\emph{ \footnotesize \textbf{Tabla 2.} Condiciones impuestas a los parámetros que se deducen de las reglas de Hume-Rothery y la termodinámica. (*) \cite{sahlberg2016}, (**) \cite{wang2015}, (***) \cite{liang2017}. }     
\end{center}   

El parámetro $\Phi$ presenta un valor crítico para $\Phi_c \approx 20$, de acuerdo con esto, las \emph{HEAs} se existen para $\Phi > \Phi_c$, a  su vez presenta múltiples estructuras cristalinas cuando $\Phi < \Phi_c$, de igual forma las \emph{HEAs} existen cuando $\Lambda > 0.95 $, componentes intermetálicos son encontrados para $\Lambda < 0.24 $. Debido a que ni la entropía de mezcla (ecuación 8) ni la entalpía de formación (ecuación 9) son buenos parámetros para describir la estabilidad de la estructura cristalina de las \emph{HEAs}, se propuso el parámetro combinado $\Omega$ para predecir la formación de SSs \cite{yang2012prediction}.    

\subsection{Energía potencial configuracional elástica de las \emph{HEAs}} 
   
Para entender que es la energía potencial elástica configuracional (o simplemente energía potencial elástica, cuyas unidades son de presión y no propiamente son de energía) tenemos que tomar como punto de partida una SS binaria y la idea se generaliza para múltiples componentes. Sí consideramos una sustancia de una sola especie atómica (visualizada como una gelatina transparente \cite{eshelby1956} para efectos prácticos) sin tener en cuenta los defectos propios de la sustancia, entonces podemos dividir el espacio en pequeños cubos (figura 6b, izquierda), en dos dimensiones el espacio seria una cuadricula (coordenadas rectangulares) como se muestra en la figura 6a (izquierda). Sí se introduce una especie atómica diferente (se introduce un defecto) el espacio seria perturbado o tensionado  (figura 6a, derecha, coordenadas curvilíneas) y cada cubo sería deformado (figura 6b, derecha) debido a que la nueva especie atómica, posee un radio atómico diferente y la entalpía de formación respecto a cada uno de las componentes atómicas es diferente, haciendo que la red o el medio se tensione o desajuste, así cada punto del espacio (desajustado) podría asociarse a un campo llamado campo elástico.   

\newpage
\begin{figure}[!htb] 
\centering
\includegraphics[width=0.6\textwidth]{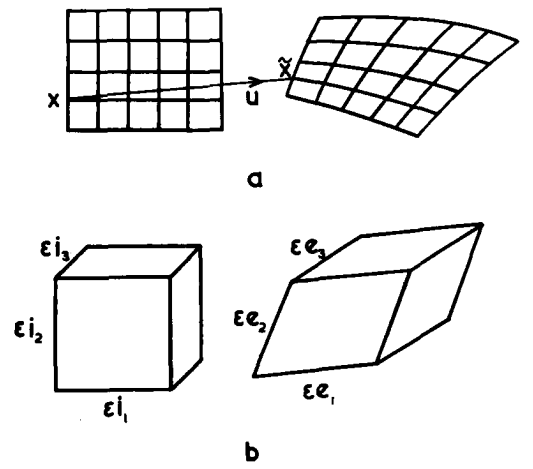}
\end{figure}
 
\begin{center}
\small
\emph{\footnotesize\textbf{Figura 6.} Deformación finita: a. Del espacio visto en dos dimensiones, b. De un cubo tridimensional del cual se compone el espacio. El vector \textbf{ u } junta un punto $x_i$ de la red sin deformar (izquierda) hasta un punto $\tilde{x}_i$ de la red deformada (derecha), siendo $\tilde{x}_1=x_1$, $\tilde{x}_2=x_2$, $\tilde{x}_3=x_3$, evidentemente el desplazamiento finito sufrido por la partícula de material originalmente en $x_i$ .Tomado de Eshelby \cite{eshelby1956}. }     
\end{center}  

Eshelby mostró \cite{kamachali2022} que la energía potencial elástica configuracional de una SS binaria, con $i-soluto$ y $h-solvente$ es,

\begin{equation} 
  e=\frac{\gamma_h(\gamma_h-1)}{\gamma_i(\gamma_i-1)}q_{ih}\lambda_{ih}^2 c_i - q_{ih}\lambda_{ih}^2 c_i^2,
 \end{equation}

donde $q_{ih}=\frac{9K_i(\gamma_h-1)}{2\gamma_h}$,  $ \gamma_{i/h}=\frac{3K_{i/h} + 4\mu_h }{3K_{i/h}}$, $K_i$ y $\mu_i$ son los módulos de compresibilidad y cizalladura de la sustancia $i$, respectivamente. 

La ecuación (16) está basada en el modelo de esfera en el agujero considerando una SS en el limite de elasticidad lineal en el nivel continuo. De esta manera Eshelby consideró la sustitución volumétrica hecha por un átomo $i$ dentro un agujero generado por la remoción de un átomo $h$ \cite{kamachali2022}. 
 
De la ecuación (16) se tiene que el primer termino es la (auto) energía debida a la propia sustitución y el segundo termino surge de la interacción elástica entre los átomos de soluto $i$. además se tiene que, $ q_{ih}\lambda_{ih}^2 = \frac{K_i \Delta V_i \Delta V_i^I}{2 V_h^2}$, en el cual $V_h$ es el volumen atómico del solvente $h$.
 
Definimos el cambio volumen total \cite[página 9]{kamachali2022}, debida a la inclusión de un átomo de soluto $i$ \cite[página 115]{eshelby1956} $\Delta V_i=\Delta V_i^\infty + \Delta V_i^I$, donde $ \Delta V_i^\infty$ es el cambio de volumen cuando la matriz es infinita (ver figura 7) y $\Delta V_i^I $ es la corrección debida a las superficies libres de otros átomos de soluto insertados y sirve para que se cumplan las condiciones de frontera. 

Las siguientes igualdades son válidas,

\begin{equation} 
 \Delta V_i=3V_h \lambda_{ih},
 \end{equation}
 
 \begin{equation} 
 \Delta V_i^\infty=\Delta V_h =\frac{3V_h \lambda_{ih}}{\gamma_h},
 \end{equation}
   
 \begin{equation} 
 \Delta V_i^I=3V_h \lambda_{ih} \frac{\gamma_h-1}{\gamma_h}.
 \end{equation} 
 
 Para encontrar la energía potencial elástica de las \emph{HEAs} el modelo de esfera en el agujero se debe generalizar para tener en cuenta la interacción entre solutos diferentes, para lograrlo Kamachali y Wang \cite{kamachali2022} consideraron la sustitución binaria $i \rightarrow j$ mostrada en la figura (7). 
 
 \begin{center} 
\includegraphics[width=5.8cm, height=9.8cm]{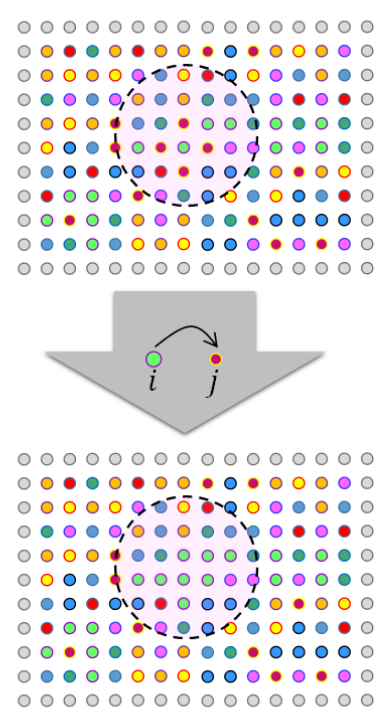} 
\small

\emph{\footnotesize\textbf{Figura 7.} El escenario de una sustitución binaria $i \rightarrow j$ dentro de un pequeño volumen de descomposición (representado por el circulo punteado), dentro de una matriz infinita. Tomado de Kamachali y Wang \cite{kamachali2022}}.   
    
\end{center}   
 
Para determinar la energía entre los defectos (átomos de la misma especie) primero se determina la interacción de un defecto con el campo elástico generado por otro. En el nivel continuo esto debe ser igual al trabajo hecho por la presión $p$ debida a los átomos de soluto ya existentes sobre los recién añadidos. A medida que se da la formación de la mezcla, la energía de interacción crece progresivamente asumiendo que  la distribución de los solutos es homogénea y al azar.  

Para un átomo de soluto añadido, en la vecindad de este, la presión es igual a,

 \begin{equation}  
 p_{1i}= -\frac{K_i \Delta V_h^I}{N_0 V_h}=-\frac{2}{3} \frac{q_{ih}\lambda_{ih}}{3N_0}.
 \end{equation} 
 
 Sí sumamos todas las interacciones de energía adicionales $i-i$ debido a la adición de $n_i$ átomos de soluto, encontramos la energía total de interacción entre todos los pares de soluto $i$, 
 
  \begin{equation}   
 e_{ii}^{int}=\frac{1}{N_0V_h}\sum_{k=1}^{n_i}P_k \Delta V_i=\frac{1}{N_0V_h}\sum_{k=1}^{n_i}kp_{1i}\Delta V_i=\frac{p_{1i}\Delta V_i}{N_0 V_h} \frac{n_i^2}{2},
 \end{equation} 
  donde se uso $\sum_{k=1}^{n}k=\frac{n(n+1)}{2}\approx n^2 .$ Reemplazando las ecuaciones (17) y (20) en (21), se tiene que, 
 
   \begin{equation}   
 e_{ii}^{int}=-q_{ih}\lambda_{ih}^2 c_i^2,
 \end{equation}
 siendo este el segundo miembro de la ecuación (16).
  
 Sí adherimos un átomo de soluto $j$, entonces la interacción entre los $n$ átomos $i$ existentes y el átomo nuevo $j$, incrementa la energía del sistema de la siguiente forma,
 
   \begin{equation}   
 p_{ni}\Delta V_j=-2c_iq_{ih}V_h\lambda_{ih} \lambda_{jh},
 \end{equation}
 y la presión del sistema incrementa como,
 
 \begin{equation}   
 p_{ni} + p_{1j}=n_ip_{1i} + p_{1j}=-\frac{2 q_{ih}\lambda_{ih}n_i}{3N_0} -\frac{2 q_{jh}\lambda_{jh}}{3N_0}, 
 \end{equation}   
  donde se usó la ecuación (20), $p_{ni}$ es la presión debida a la interacción $i-i$ entre todos los pares de átomos de soluto $i$, y $p_{1j}$, da la presión introducida por el nuevo átomo $j$.  
 
 Se puede construir de manera análoga a la ecuación (22) en el caso que la sustitución sea $j\rightarrow i$, los términos de interacción $j-j$. 
 
 Para los términos de interacción $i-j$ se tiene que, 
   
 \begin{equation}   
 e_{ij}^{int}=\frac{p_{ni} n_j \Delta V_j}{N_0 V_h}=-2q_{ih}\lambda_{ih} \lambda_{jh} c_i c_j.
 \end{equation} 
   
  Si intercambiamos la sustitución $i \rightarrow j$ por una $j \rightarrow i$, entonces $e_{ij}^{int}=e_{ji}^{int}$, esto se puede expresar estadísticamente como $e_{ij}^{int}=-2\sqrt{q_{ih}q_{jh}}\lambda_{ih} \lambda_{jh} c_i c_j $.
  
 Generalizando este proceso se obtiene la energía potencial elástica de formación de una solución solida de $N$ componentes que obedece la sustitución $i \rightarrow j$ entre todos los pares de combinaciones posibles,  
 
 \begin{equation}     
 e= \sum_{\alpha\neq h}^{N}q_{\alpha h}\lambda_{\alpha h}^2  \frac{\gamma_h(\gamma_h-1)}{\gamma_\alpha(\gamma_\alpha-1)} c_\alpha - \sum_{\alpha \neq h}^{N}q_{\alpha h}\lambda_{\alpha h}^2 c_\alpha^2 - 2\sum_{\scriptsize \begin{matrix}
 \alpha,\beta \neq h \\        
  \alpha\neq \beta 
\end{matrix}}^{N}\sqrt{q_{\alpha h} q_{\beta h}}\lambda_{\alpha h} \lambda_{\beta h} c_{\alpha} c_{\beta}.    
 \end{equation}

 Para $\alpha,\beta-solutos $ y $h-solvente$;  donde $\alpha=i, j, ..., N$; $\beta=i, j, ..., N$. Por ejemplo para la $HEA$ $TiVZrNbHf$, se tiene que $N=5$. El solvente $h$ puede ser cualquiera de sus componentes, esto deja invariable la energía elástica de formación $e$ \cite[página 1]{kamachali2022} medida en unidades de presión $(GPa)$; esto es así debido a que los átomos en este capítulo son considerados como esferas de billar o canicas sólidas de diferentes tamaños, haciendo que el concepto de solvente siga siendo el que formalmente se encuentra en la literatura.

En la ecuación (26) el primer término es debido a la autoenergía de los átomos de soluto añadidos, (ver primer término de la ecuación 16), el segundo término (ver segundo término de la ecuación 16, ver ecuación 22)  es la energía de interacción entre pares de átomos ($\alpha, \alpha$), el tercer término es la energía de interacción entre pares de átomos  distintos ($\alpha, \beta$).  La ecuación (26) incluye efectos anisotrópicos, pero estos pueden ser omitidos si consideramos que las \emph{HEAs} son elásticamente isotrópicas, asumiendo que estas poseen un coeficiente de compresibilidad equivalente para todas las sustancias,  $ K_\alpha \equiv K_h $, con lo cual $\gamma_\alpha=\gamma_h$.  

De acuerdo con Kamachali se tiene que \cite{kamachali2022},        

\begin{equation}
 q_{\alpha h}\lambda_{\alpha h}^2 = \frac{K_\alpha \Delta V_\alpha \Delta V_\alpha^I}{2 V_h^2}=\frac{9}{2}K_\alpha \lambda_{\alpha h}^2 \frac{\gamma_h-1 }{\gamma_h },
\end{equation}
aquí se usó las ecuaciones (18) y (19) (con la notación adecuada). De la ecuación (27) se deduce la siguiente igualdad,

\begin{equation}
 q_{\alpha h} =\frac{9K_\alpha (\gamma_h-1)}{2\gamma_h}, 
\end{equation}
análogamente,  

\begin{equation}
 q_{\beta h} =\frac{9K_\beta (\gamma_h-1)}{2\gamma_h},
\end{equation}
reemplazando las ecuaciones (27), (28) y (29) en la ecuación (26) y simplificando se obtiene, 
    
\begin{equation}
 e = \frac{ 9(\gamma_h-1)}{ \gamma_h} \left [   \sum_{\alpha\neq h}^{N} \frac{K_\alpha \lambda_{\alpha h}^2}{2} \left ( \frac{\gamma_h(\gamma_h-1)}{\gamma_\alpha(\gamma_\alpha -1)}c_\alpha- c_\alpha^2 \right )- \sum_{\scriptsize \begin{matrix}
 \alpha,\beta \neq h \\       
  \alpha\neq \beta 
\end{matrix}}^{N}\sqrt{K_{\alpha} K_{\beta }}\lambda_{\alpha h} \lambda_{\beta h} c_{\alpha} c_{\beta} \right ],   
\end{equation}
donde,  

\begin{equation}
 \gamma_{\alpha/h}=\frac{3K_{\alpha/h} + 4\mu_h }{3K_{\alpha/h}}, 
\end{equation}
de aquí se deduce la siguiente relación,

\begin{equation}
 \frac{\gamma_h}{\gamma_\alpha}=\frac{K_\alpha}{K_h} \left ( \frac{3K_h+4\mu_h}{3K_\alpha+4\mu_h}  \right ),
\end{equation}
de la ecuación (31) es evidente que, 

\begin{equation}
 \gamma_{\alpha/h}-1=\frac{ 4\mu_h }{3K_{\alpha/h}},
\end{equation} 
de lo cual se obtuvo, 

\begin{equation}
 \frac{\gamma_{h}-1}{\gamma_{\alpha}-1}=\frac{ K_\alpha }{K_h}.
\end{equation} 

Sí asumimos la existencia de constantes elásticas isotrópicas, $\gamma_\alpha = \gamma_h $ y $ K_\alpha = K_h$,  las ecuaciones (32) y (34) se reducen a la unidad, 

\begin{equation}
 \frac{\gamma_h}{\gamma_\alpha}=1,
\end{equation}

\begin{equation}
 \frac{\gamma_h-1}{\gamma_\alpha-1}=1.
\end{equation} 
  
  Sustituyendo las ecuaciones (33), (35) y (36), en la ecuación (30), se llegó a lo siguiente, 
  
  \begin{equation} 
 e = \frac{ 12\mu_h }{ \gamma_h}   \sum_{\alpha\neq h}^{N} \frac{\lambda_{\alpha h}^2}{2} \left ( c_\alpha- c_\alpha^2 \right )- \frac{ 12\mu_h }{ \gamma_h}   \sum_{\scriptsize \begin{matrix}
 \alpha,\beta \neq h \\       
  \alpha\neq \beta 
\end{matrix}}^{N} \lambda_{\alpha h} \lambda_{\beta h} c_{\alpha} c_{\beta}.   
\end{equation}

De acuerdo con Eshelby \cite[Página 108, ecuación 8.7]{eshelby1956},

\begin{equation}
 \gamma_h=\frac{3K_h+4\mu_h}{3K_h}=3\frac{1-\nu_h}{1+\nu_h},
\end{equation}  
     donde $\nu_h $ es el cociente de Poisson de un átomo de solvente $h$, por lo tanto, 
     
\begin{equation}
\frac{ 12\mu_h }{ \gamma_h}=4\mu_h \left (\frac{1+\nu_h}{1-\nu_h} \right ) ,
\end{equation}
sustituyendo la ecuación (39) en la ecuación (37) se obtiene lo siguiente, 

  \begin{equation} 
 e = q   \sum_{\alpha\neq h}^{N} \lambda_{\alpha h}^2 \left ( c_\alpha- c_\alpha^2 \right )- 2q   \sum_{\scriptsize \begin{matrix}
 \alpha,\beta \neq h \\       
  \alpha\neq \beta 
\end{matrix}}^{N} \lambda_{\alpha h} \lambda_{\beta h} c_{\alpha} c_{\beta},   
\end{equation}
     donde, $q=2\mu_h \left (\frac{1+\nu_h}{1-\nu_h} \right )$, es una constante que nació de la condición de isotropía  ($\gamma_\alpha = \gamma_h $, $ K_\alpha = K_h$).  La elección del  solvente $h$ es arbitraria, los coeficientes $\lambda_{\alpha h}$ son antisemíticos $\lambda_{\alpha h}=-\lambda_{h \alpha}$, y cumplen $\lambda_{\alpha h}=\lambda_{\alpha k}+\lambda_{kh}$, además debido a la composición equimolar de las \emph{HEAs} no se distingue entre solvente (ver figura 1b), así que es conveniente reemplazar $\mu_h$ por $\bar{\mu}$ y $\nu_h$ por $\bar{\nu}$, donde  $\bar{\mu}=\sum_{k=1}^{N}c_k \mu_k$  y $\bar{\nu}=\sum_{k=1}^{N}c_k \nu_k$, donde $ \mu_k$ y $\nu_k$ son los coeficientes de cizalladura y el cociente de Poisson de la $k$ ésima sustancia respectivamente, con lo cual se asumió en este trabajo que, $q=2\bar{\mu} \left (\frac{1+\bar{\nu}}{1-\bar{\nu}} \right )$. De acuerdo con Eshelby estas derivaciones están en el limite de elasticidad lineal asumiendo una SS aleatorizada \cite{kamachali2022}, ver figura 1b. Es posible simplificar la ecuación (40) sustituyendo la definición del parámetro $\delta$ ( ver la ecuación 5) en ella (ver apéndice A), esto es,  

\begin{equation}
    e=q\delta^2. 
\end{equation}
      
    Para este trabajo se usó la energía elástica configuracional $e$ $(GPa)$ dada por la ecuación (41) como si fuese una constante que actuará sobre la energía de cada átomo, esta constante está directamente relacionada a la energía por partícula explicada más adelante en el capítulo (4.4), comparar con las ecuaciones (26 y 30) para el cual $\gamma_\alpha \neq \gamma_h $ y $ K_\alpha \neq K_h$.

\subsection{Exceso de entropía de mezcla}

En el proceso de fabricación de \emph{HEAs} cuando esta se encuentra a la temperatura de fundición, antes de que  solidifique como SS, en este estado la entropía de mezcla viene descrita por,
\begin{equation} 
     S_{mez}=-k_B\sum_{i=1}^{N} c_i ln(c_i) 
\end{equation}  
 y esta tiende a la entropía de formación ideal $ S_{id}=k_B\sum_{i=1}^{N} c_i ln(N)$, donde  $N$ es el número de especies atómicas diferentes. Esto es así debido a que el agitamiento térmico es tan fuerte que prácticamente las interacciones interatómicas no tienen relevancia, una vez la \emph{HEA} empieza a solidificarse, el agitamiento térmico ya no es tan fuerte (disminuye su frecuencia) y las interacciones interatómicas juegan un papel importante, en este punto la entropía de la \emph{HEA} la llamamos entropía corregida o correlacionada $S_{corr}$, definida de acuerdo con Ding \cite{ding2018} como, 

\begin{equation} 
    S_{corr}=S_{id}+S_E,
\end{equation} 
 donde $S_E$ ($\leq 0$) denota el exceso de entropía de mezcla, y tiene el efecto de disminuir $S_{id}$  cuando la \emph{HEA} se encuentra por debajo de la temperatura de fundición y la variedad de interacciones interatómicas cobra importancia frente a la entropía configuracional (ecuación 41). $S_E$ se define de acuerdo con He \cite{ding2018}, como muestra la siguiente relación, 

\begin{equation}  
    S_E=k_B \left [ 1 + \frac{x}{2}-ln(x)+ln(1-e^{-x}) - \frac{x}{2} \cdot \frac{1+e^{-x}}{1-e^{-x}} \right],
\end{equation}
donde el parámetro adimensional $x=\frac{\Delta \epsilon}{k_B T}$ se define para la fluctuación de la energía normalizada $\Delta \epsilon$,  y se describe de acuerdo con Ding \cite{ding2018}, como,  

\begin{equation}
    x=x_e + x_c,  
\end{equation}
siendo $x_e$ la contribución a $S_E$ debido a la diferencia de tamaños atómicos, y $x_c$ es la contribución a  $S_E$ debido a los diferentes enlaces químicos que se pueden dar y están limitados por la entalpía de formación entre pares de especies atómicas diferentes tomados a partir del modelo de Miedema \cite{takeuchi2005classification}. De acuerdo con Ding \cite{ding2018} $x_\epsilon$  es equivalente a la siguiente ecuación, 

\begin{equation}
    x_e=4.12 \delta \sqrt{\frac{\Bar{K} \Bar{V}}{k_B T}},
\end{equation}
 siendo $\Bar{K}$ y $\Bar{V}$, el módulo de compresibilidad promedio y el volumen atómico promedio, respectivamente, $k_B$ es la constante de Boltzmann. Reemplazando la definición dada en la ecuación (41) $e=q\delta^2$, se tiene que,
 
\begin{equation}
    x_e=4.12 \sqrt{\frac{e\Bar{K} \Bar{V} }{q k_B T}}.
\end{equation}
De manera análoga a la ecuación (38) se representó el módulo de compresibilidad promedio $\Bar{K}$, por medio de la siguiente relación, 
\begin{equation}
    \frac{3\Bar{K} + 4\Bar{\mu}}{3\Bar{K}}=3 \frac{1-\Bar{\nu}}{1+\Bar{\nu}},
\end{equation}
donde $\Bar{\nu}$ es el cociente de Poisson promedio, sí,  
\begin{equation}
    q=2\Bar{\mu} \left ( \frac{1+\Bar{\nu}}{1-\Bar{\Bar{\nu}}}\right ),
\end{equation}
 podemos relacionar directamente $\Bar{K}/q$, quedando, 
\begin{equation}
    \Bar{s}=\frac{\bar{K}}{q}=\frac{2}{3}\cdot \frac{1}{3-\frac{1}{\frac{1-\bar{\nu}}{1+\bar{\nu}}}},
\end{equation}
con lo cual reescribimos la ecuación (47),
\begin{equation}
    x_e=4.12 \sqrt{\frac{e\Bar{s} \Bar{V} }{ k_B T}},
\end{equation}
el volumen atómico promedio se definió como, $\bar{V}=\sum_{k=1}^{N}c_k V_k $, con $V_k$ el volumen atómico de la especie atómica $k$. De está manera se escribió la contribución a la entropía corregida  $S_{corr}$ (ecuación 43) debida a la energía elástica configuracional $e$. De acuerdo con Ding \cite{ding2018} $x_c$  es equivalente a la siguiente ecuación,

\begin{equation}
    x_c=2 \sqrt{\frac{\sqrt{\sum_{i}^{ } \sum_{
 j,  i \neq j}^{} c_i c_j \left (  H_{i j} - \Bar{H} \right )^2 }}{k_B T}},
\end{equation}

  la temperatura es tal que $T<T_f$, $H_{i j}$ es la entalpía de formación entre la especie atómica $i$ y $j$, calculado a partir del modelo de Miedema \cite{takeuchi2005classification, ding2018,he2016configurational}, $\Bar{H}$ es el promedio de $H_{i j}$. De acuerdo con He \cite{he2016configurational} las \emph{HEAs}  están definidas en el rango $0.85 < S_{corr}/S_{id}<1$, para múltiples estructura cristalinas están en el rango $0.7 < S_{corr}/S_{id}<0.85$, y aleaciones amorfas en el rango $0.4 < S_{corr}/S_{id}<0.6$.

\subsection{Modelo para la entalpía de Mezcla $\Delta H_{mez}^*$ y entalpía corregida $\Delta H_{corr}$}  

Dado que el parámetro $\Omega$ (ecuación 12)  se usó porque ni la entalpía de mezcla ni la entropía de mezcla son buenos describiendo la estabilidad de la \emph{HEA} \cite{yang2012prediction}, se propuso en este trabajo dar una definición alterna a la entalpía de formación; para lograrlo se usaron dos definiciones diferentes de la función de partición \emph{Z}, la primera se tomó de Takeuchi e Inoue \cite[ecuación 14]{takeuchi2001quantitative}, y se define así,

\begin{equation}
    Z=e^{\frac{\Delta G}{RT}},
\end{equation}
 sustituyendo $\Delta G$ por $\Delta G_{mez}$ (ecuación 11) y $T$ por $T_f$ (ecuación 10), simplificando un poco se encontró el siguiente modelo para la entalpía de formación,   

\begin{equation}
    \Delta H_{mez}^*=RT_f ln(Z) - k_BT_f \sum_{i = 1}^{N} c_i ln(c_i).
\end{equation}

La ecuación (54) es similar en estructura a la ecuación de Bragg-Williams para la energía libre de Gibbs \cite{morral2017regular}. De forma análoga a la ecuación (54) puede ser hallada la entalpía corregida $\Delta H_{corr}$, 

\begin{equation}
    \Delta H_{corr}=k_BT ln(Z) - k_BT \sum_{i = 1}^{N} c_i ln(c_i),
\end{equation}
para $T\leq T_f$ una vez que la \emph{HEA} ya se ha fabricado. 

La segunda definición de $Z$ de acuerdo con la mecánica estadística,  puede ser expresada con ayuda de la función de partición canónica \cite{he2016configurational} aplicada a sistemas cuánticos o discretos, definida como,  
 
 \begin{equation}
Z=Z_{N^*}=\sum_{i}^{\Omega^*}e^{-\frac{E_i}{k_B T}}, 
\end{equation}
 $\Omega^*$ es el número de microestados disponibles, $E_i$ es la energía del microestado de un sistema de $N^*=nN_A$ partículas, donde $n$ es el número de moles; este microestado queda identificado mediante una configuración que especifica la ocupación de cada partícula de los estados de un cuerpo, como,
 \begin{equation}
i=\left \{k_1,k_2,...,k_{N^*}\right \}, 
\end{equation}

 \begin{equation}
E_i=\epsilon_{k1} + \epsilon_{k2} +...+\epsilon_{kN^*}, 
\end{equation}
 donde, $\epsilon_{k1}$ es la energía del primer átomo (tomado al azar) en el estado $k$, $\epsilon_{k2}$ es la energía del segundo átomo (tomado al azar) en el estado $k$, de esta manera se continua hasta llegar a la contribución del último átomo ($\epsilon_{kN^*}$) al estado $k$,  así con la ayuda de las ecuaciones (57) y (58) reescribimos $Z$,
\begin{equation}
Z=\sum_{k_1,k_2,...,k_{N^*}}^{\Omega^* } e^{-\frac{\epsilon_{k1} + \epsilon_{k2} +...+\epsilon_{kN^*}}{k_B T}}. 
\end{equation}

Sí las partículas son distinguibles, la ecuación (59) se transforma en,

\begin{equation}
Z=\left ( \sum_{k_1}^{\Omega^* } e^{-\frac{\epsilon_{k1}}{k_B T}} \right ) \left ( \sum_{k_2}^{\Omega^* } e^{-\frac{\epsilon_{k1}}{k_B T}} \right ) ... \left ( \sum_{kN^*}^{\Omega^* } e^{-\frac{\epsilon_{kN^*}}{k_B T}} \right ). 
\end{equation}

A continuación consideremos los posibles microestados asociados al sistema. De acuerdo con el concepto de $HEA$, esta es una SS mono cristalina donde las diferentes especies atómicas, en su mayoría metales de transición, se encuentran entre el $5\%$ y el $35\%$. Sí consideramos dos aportes principales a la energía del microestado $E_i$, los cuales se deducen \emph{primero} de la exigencia, de que las componentes atómicas se encuentren ubicadas en el espacio lo más aleatoriamente posible y \emph{segundo} del concepto de energía potencial elástica configuracional (esta energía es equivalente a tener todos los átomos separados y luego juntarlos en posiciones aleatorias, ejerciendo una fuerza externa sobre cada uno de ellos, equivalente al trabajo necesario para ubicarlos en dichas posiciones)\cite{kamachali2022}. Ambas exigencias sugieren que los  microestados asequibles a los átomos, son aquellos que maximicen la energía (recuerde que las \emph{HEAs} se fabrican alcanzando su punto de fundición). 

Justificado por ambas exigencias, en la figura (8) se elaboró un sencillo modelo para los microestados de energía, suponiendo que, cada átomo solo puede estar en
presencia de $N - 1$ átomos, ya sean diferentes o no, esto se representa como circunferencias (de áreas exageradas), cada una de estas circunferencias, en las cuales los
átomos solo son distinguibles por sus tamaños y no por sus posiciones, se repite abarcando
todo el contenido macroscópico, los círculos grises indican que la matriz se extiende al
infinito. A continuación supongamos que existen muchos microestados posibles (del orden del número de átomos) de entre los cuales destacan dos microestados, uno en el que todos los átomos se encuentran a mínima energía (figura 8b) debido a que todos los átomos son iguales en cada circunferencia mostrada en la figura, indicando que no hay energía suficiente para la máxima aleatoriedad y la red no se ha desajustado localmente, y otro para el que todos los átomos están a su máxima energía (figura 8a) producto de la máxima aleatoriedad posible (todos los átomos son diferentes en cada circunferencia mostrada en la figura)  y el máximo desajuste de la red que contribuye a la energía  elástica configuracional.

 \begin{center} 
\includegraphics[width=12.5cm, height=12cm]{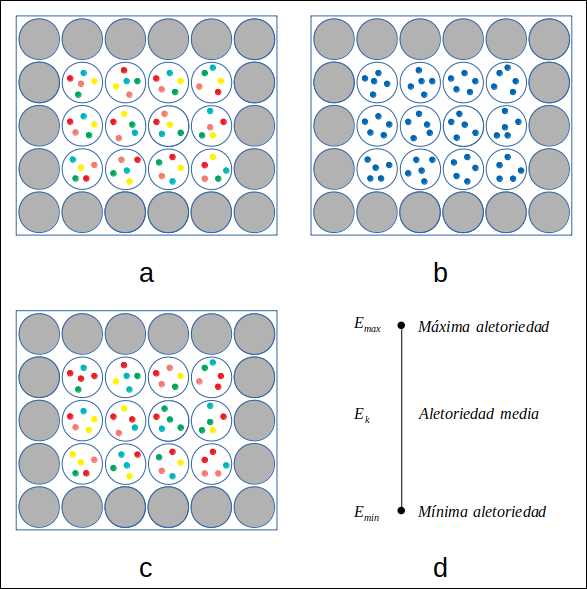} 

\small

\emph{ \footnotesize 
\textbf{Figura 8.} Microestados asequibles al sistema, para una \emph{HEA} de cinco ($N=5$)  componentes atómicas (puntos de colores). \textbf{a}. Microestado con energía máxima $E_{max}$, \textbf{b.} Microestado con mínima energía $E_{min}$. \textbf{c.} La energía de los microestados se ubica en el rango, $E_{min} < E_k < E_{max}$, debido a que ese microestado no se encuentra en la máxima (o mínima) aleatoriedad, \textbf{d.} Comparación energética del microestado de máxima energía $E_{max}$ (figura a), mínima energía $E_{min}$ (figura b) y el de energía intermedia (figura c). }    
\end{center} 

De acuerdo con la figura (8), las \emph{HEAs} estarían bien definidas por la figura (a), la cual representa los microestados de máxima energía, esto indica que en el proceso de fabricación, estas tienden a un solo microestado ($\Omega^*=1$, ver ecuación 55), el microestado de máxima energía $E_{max}$ cuando existe la presencia de cinco átomos diferentes, ver figura (8), de esto se deduce de acuerdo con la ecuación (60) que,

\begin{equation}
Z=\left ( \sum_{k_1}^{1 } e^{-\frac{\epsilon_{k1}}{k_B T}} \right ) \left ( \sum_{k_2}^{1 } e^{-\frac{\epsilon_{k2}}{k_B T}} \right ) ... \left ( \sum_{kN^*}^{1 } e^{-\frac{\epsilon_{kN^*}}{k_B T}} \right ), 
\end{equation}
como los microestados se encuentran en la máxima energía $E_{max}$, como se dijo, estos microestados solo serán de energía máxima, si para cada microestado el numero $N$ de átomos es diferente (ver figura 8), es decir,

\begin{equation}
    \epsilon_{k1} \neq \epsilon_{k2} \neq ... \neq \epsilon_{kN},
\end{equation} 
estos microestados se repiten por toda la \emph{HEA} (ver figura 8a), así que es equivalente a multiplicar el número total de átomos iguales por la función de partición de ese átomo en específico, se deduce de la ecuación (56) que, 

\begin{equation}
Z=\left ( c_1 N^* \sum_{k_1}^{1 } e^{-\frac{\epsilon_{k1}}{k_B T}} \right ) \left ( c_2 N^*\sum_{k_2}^{1 } e^{-\frac{\epsilon_{k2}}{k_B T}} \right ) ... \left ( c_N N^*\sum_{kN}^{1 } e^{-\frac{\epsilon_{kN}}{k_B T}} \right ), 
\end{equation}

\begin{equation}
Z=\left ( c_1 N^*  e^{-\frac{\epsilon_{max1}}{k_B T}} \right ) \left ( c_2 N^* e^{-\frac{\epsilon_{max2}}{k_B T}} \right ) ... \left ( c_N N^* e^{-\frac{\epsilon_{maxN}}{k_B T}} \right ), 
\end{equation}

\begin{equation}
Z=N^{*N}c_1c_2...c_N e^{-\frac{E_{max}}{k_B T}}, 
\end{equation}
donde $E_{max}=\epsilon_{max1}+\epsilon_{max2}+...+\epsilon_{maxN}$, siendo \emph{N} el número de componentes de la \emph{HEA} y $N^*$ el número total de átomos, $c_j$ es la concentración molar de la especie atómica $j$, y cumplen la condición de normalización, $\sum_{j}^{N}c_j=\sum_{k}^{N} \frac{n_k N_A}{N^*}=\sum_{k}^{N}\frac{n_k}{n}=1$.
 
Como se discutió arriba, las \emph{HEAs} se fabrican buscando que cada átomo se encuentre lo más distribuido al azar, esto se logra en parte alcanzado la temperatura de fundición $T_f$ (ecuación 10) de cada especie atómica, está energía (-$T_f\Delta S_{mez}$) junto con la entalpía ($\Bar{H}$) cuando esta es negativa, entran en competición con la energía potencial elástica configuracional $e$ (ecuación 41) [Kamachali and Wang, 2022]. 

Motivado por las ecuaciones (51) y (52), se asumió en este trabajo que, la energía que experimenta cada átomo cuando la \emph{HEA} se ha fabricado, se define como, 

\begin{equation}
 \epsilon_{i}=\epsilon_{maxi}= \begin{cases}
 & s_i V_i e,  \text{ sí } \bar{H} \geq 0 \\
 & s_i V_{i} e + \sqrt{\sum_{\scriptsize \begin{matrix} j=1, j\neq i\end{matrix}}^{N} \left (  H_{i j} - \Bar{H} \right )^2}, \text{ sí } \bar{H} < 0,
\end{cases}
\end{equation}
 donde $s_i=\frac{K_i}{q}=\frac{2}{3}\cdot \frac{1}{3-\frac{1}{\frac{1-\nu_i}{1+\nu_i}}}$, $V_i$ es volumen atómico de la especie atómica $i$. Siendo $s_iV_ie$ la contribución a la energía del microestado $ E_i$ debido a la energía potencial elástica configuracional $e$, tomada como una constante en esta parte del trabajo, el término con raíz es la contribución a $\epsilon_i $ gracias a la entalpía de formación que se da en presencia de las diferentes interacciones interatómicas y es $\epsilon_i = s_i V_{i} e$, sí las interacciones interatómicas en promedio son endotérmicas ($\bar{H}\geq 0$), o es $\epsilon_i < s_i V_{i} e$, sí las interacciones interatómicas en promedio son exotérmicas ($\bar{H}<0$) \cite{kamachali2022}. 

Reemplazando la ecuación (65) en la definición dada por la ecuación (54), se redefine $\Delta H_{mez}^*$,

\begin{equation}
    \Delta H_{mez}^*=k_BT_f ln\left ( N^{*N} c_1...c_N\right )- E_{max} - k_B T_f \sum_{i=1}^{N}c_i ln(c_i),
\end{equation}
si tenemos en cuenta que, $\sum_{i=1}^{N}c_i ln(c_i)=ln\left ( c_1^{c_1}...c_N^{c_N}  \right )$, entonces se tiene que,

\begin{equation}
 \Delta H_{mez}^*=k_B T_f ln\left (N^{*N} c_1^{1-c_1}...c_N^{1-c_N}  \right )  - E_{max}  
\end{equation}
donde $E_{max}$ es el microestado de máxima energía ($E_{max}=\epsilon_{max1}+...+\epsilon_{maxN}$), se ha usado $\Delta H_{mez}^*$ para diferenciarla de $\Delta H_{mez}$ (ecuación 9). Análogamente se halló la entalpía corregida ($\Delta H_{corr}$), para una temperatura $T<T_f$ después de que la \emph{HEA} haya sido fabricada,

\begin{equation}
 \Delta H_{corr}=k_B T ln\left (N^{*N} c_1^{1-c_1}...c_N^{1-c_N}  \right )  - E_{max}.  
\end{equation}

Las ecuaciones (68) y (69) fueron los aportes hechos por este trabajo a la entalpía de formación de las \emph{HEAs}.

\newpage
\section{Metodología}  
 
Teniendo presente que para cumplir el objetivo principal de este trabajo se adoptaron las condiciones impuestas a los parámetros que se derivan de las reglas de Hume-Rothery y la termodinámica (tabla 2), se consideró aparte de estos parámetros, la ecuación (40), de la cual se pueden hacer algunas estimaciones sobre la estabilidad de las SSs \cite{kamachali2022} (consideradas elásticamente isótropas) si se tiene en cuenta que la variación en la concentración $\Delta c_i$ produce una variación en el potencial $e$ de la siguiente manera (ver apéndice B), 

\begin{equation*}       
 \left.\frac{\Delta e \left (  c_i, c_j  \right )}{\Delta c_i}  \right |_j = q \left ( \lambda_{ih}^{2} - \lambda_{jh}^{2} \right )+2q \left (c_j \lambda_{jh}^{2} - c_i \lambda_{ih}^{2} - \lambda_{ih} \lambda_{jh} \left ( c_j - c_i \right ) \right )  
 \end{equation*} 
 \begin{equation} 
  \hspace{-3cm}- q\left ( \lambda_{ih}-  \lambda_{jh}  \right )^{2} \Delta c_i,  
 \end{equation}
donde la barra con el subíndice $j$ denota la sustitución $i \rightarrow j$ , análogamente a partir de la ecuación (40), podemos conocer como es la variación del potencial respecto a la variación de la concentración $\Delta c_j $, con la sustitución binaria $j \rightarrow i $, resultando,  

 \begin{equation*}       
 \left.\frac{\Delta e \left (  c_i, c_j  \right )}{\Delta c_j}  \right |_i = -q \left ( \lambda_{ih}^{2} - \lambda_{jh}^{2} \right )-2q \left (c_j \lambda_{jh}^{2} - c_i \lambda_{ih}^{2} - \lambda_{ih} \lambda_{jh} \left ( c_j - c_i \right ) \right )  
 \end{equation*}
 \begin{equation} 
   \hspace{-3cm} - q\left ( \lambda_{ih}-  \lambda_{jh}  \right )^{2} \Delta c_j.  
 \end{equation}

 Sí comparamos las ecuaciones (70) y (71) se encuentra el siguiente resultado,

\begin{equation} 
  \left.\frac{\Delta e \left (  c_i, c_j  \right )}{\Delta c_j}  \right |_i  = - \left.\frac{\Delta e \left (  c_i, c_j  \right )}{\Delta c_i}  \right |_j.  
 \end{equation}         
     
 Otro resultado interesante aparece sí en las ecuaciones (70) y (71) hacemos $\lambda_{ih} = \lambda_{jh} $, entonces,      
   
  \begin{equation} 
   \left.\frac{\Delta e \left (  c_i, c_j  \right )}{\Delta c_i}  \right |_j=0,  
 \end{equation} 
 
  \begin{equation} 
   \left.\frac{\Delta e \left (  c_i, c_j  \right )}{\Delta c_j}  \right |_i=0,  
 \end{equation}
 la variación de la energía potencial elástica desaparece respecto a cualquiera de las variaciones de las concentraciones ($i$ o $j$) , debido a que la condición $\lambda_{ih} = \lambda_{jh} $, nos indica que la red no se desajusta ante variaciones en la concentraciones $\Delta c_j=-\Delta c_i $, en cuyo caso $ r_i=r_j$, ver ecuación (2), se estaría reemplazando un átomo por otro de la misma especie (o del mismo tamaño).     

Sí se hace $ \lambda_{ih}= \lambda_{max}$, $ \lambda_{jh}= \lambda_{min}$, de tal forma que se cumpla que $ \lambda_{max}>\lambda_{min}$ y además, si $\lambda^{*}=\sqrt{\lambda_{max}^2-\lambda_{min}^2} \propto \lambda_{max}-\lambda_{min}  $, es evidente en la ecuación (70) que, 

  \begin{equation} 
   \left.\frac{\Delta e \left (  c_i, c_j  \right )}{\Delta c_j}  \right |_i\rightarrow \lambda^{*}.  
 \end{equation} 

Según Kamachali y Wang [Kamachali, 2022] se ha encontrado experimentalmente que, una condición necesaria para la formación de SSs, es que $\lambda^{*} < 0.16$, condición que se adopto en este trabajo.   
   
\subsection{Aplicación en FORTRAN90} 

Se tabularon con la ayuda de FORTRAN90 un total de 71 elementos químicos (apéndice C), de estos se tomaron las siguientes propiedades físicas y químicas,  radio atómico , electronegatividad de Pauling, temperatura de fundición, concentración de electrones de valencia, densidad numérica, masa atómica  \cite{miracle2017}, módulo de compresibilidad  \cite{makino2000estimation}, módulo de cizalladura y el cociente de Poisson \cite{samsonov1968mechanical}. El algoritmo usado fue el siguiente, 

\begin{figure}[!htb]   
\centering
\includegraphics[width=0.7\textwidth]{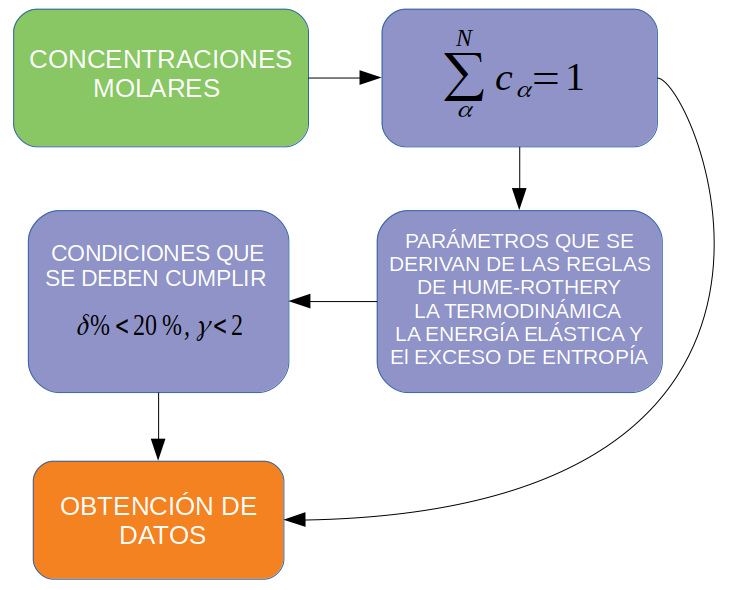} 
 
\end{figure} 

\begin{center} \small 
\emph{\footnotesize \textbf{Figura 9.} Algoritmo generalizado de la aplicación en FORTRAN90 (apéndice C). Las condiciones que se deben cumplir (cuarta etapa) fueron tomadas al azar, por encima de las condiciones  impuesta a los parámetros $\delta\%$ y $\lambda$ (tabla 2), se supuso que por debajo de estos limites los parámetros dados en la tabla (2) estarían bien definidos, los resultados mostraron que la asunción es correcta. }     
\end{center}  
   
El algoritmo mostrado en la figura (9) consta de cinco pasos o etapas, en la \emph{primera} etapa (parte superior izquierda) el programa realizó un barrido por las diferentes concentraciones molares de las diferentes especies atómicas, en la \emph{segunda} etapa se  buscaron que las sumas de las concentraciones molares sea la unidad (normalización), en la \emph{tercera} etapa se calcularon los valores de los parámetros que se deducen de las reglas de Hume-Rothery y la termodinámica (tabla 2), también se calcularon el exceso de entropía ($S_E$) y la entropía corregida o correlacionada ($S_{corr}$) (ecuación 43) buscando que se cumpliera la condición $0.85 < S_{corr}/S_{id}<1$, al igual que se calcularon los valores dados por la ecuaciones (70 y 75) para el átomo de menor radio atómico (representado por el factor $\lambda_{jh}$) y de mayor radio atómico (representado por el factor $\lambda_{ih}$), en la \emph{cuarta} etapa estos parámetros se restringen a los limites  $\delta\%<20\%$ y $\lambda<2$, se supuso que con estos limites los valores de la tabla (2) estarían bien definidos, para finalmente pasar a la \emph{quinta} y última etapa que es la obtención de datos. Aquí, se emparejaron las concentraciones molares escogidas en la segunda etapa con los parámetros calculados en la tercera etapa que a su vez fueron restringidos en la cuarta etapa. De esta manera se escribieron los datos en archivos independientes para posteriores lecturas y visualizaciones de las gráficas.

  \newpage
\section{Análisis y discusión de resultados}

De entre todas las familias de aleaciones posibles (o \emph{HEAS}), se escogieron dos familias en especial, la familia $AlCoCrFeNi$ estudiada por Liang y Schmid \cite{liang2017} y la familia $TiZrHfNbV$ estudiada por Sahlberg \cite{sahlberg2016} siendo esta última, una de las aleaciones más prometedoras en cuanto a futuras aplicaciones para el almacenamiento de hidrógeno. Estos resultados son similares a los resultados experimentales obtenidos por  Wang  para los parámetros $\delta$ y $\gamma$ \cite[figura 2]{wang2015}, y a los resultados hallados por Liang y Schmid \cite{liang2017}  para la aleación $Al_xCoCrFeNi$ con ayuda del  método de CALPHAD para los parámetros $\delta$, $\gamma$, $\Lambda$, $VEC$, $\Delta \chi$, $\Delta S_{mez}$, $\Delta H_{mez}$ y $\Omega$. Se obtuvieron un total de 238030 aleaciones con sus respectivas proporciones molares tanto como para el $AlCoCrFeNi$ como para el $TiVZrNbHf$. 
 
\subsection{Resultados que se deducen de las reglas de Hume-Rothery y la termodinámica.}
A continuación en las figuras (10 y 12) se presentan los resultados que se obtuvieron para la aleación \emph{AlCoCrFeNi}, para esta aleación se graficaron los parámetros $\delta \%, \gamma, \Lambda, \Delta \chi$. En la figura (11) se graficó el $VEC$ (ecuación 7). 

\begin{figure}[!htb]   
\centering
\includegraphics[width=0.8\textwidth]{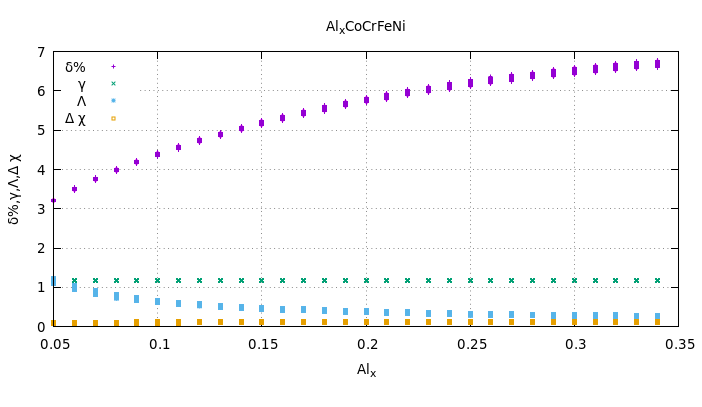} 
 
\begin{center} \small 
\emph{\footnotesize \textbf{Figura 10.} Resultado de la programación en FORTRAN90. Para la aleación $AlCoCrFeNi$ se graficaron los parámetros $\delta \%$ (ecuación 3), $\gamma$ (ecuación 6), $\Lambda$ (ecuación 15) y $\Delta \chi$ (ecuación 4). Estos resultados son positivos de acuerdo con  Liang y Schmid \cite{liang2017}}. 
\end{center} 
\end{figure}

\begin{figure}[!htb]   
\centering
\includegraphics[width=0.93\textwidth]{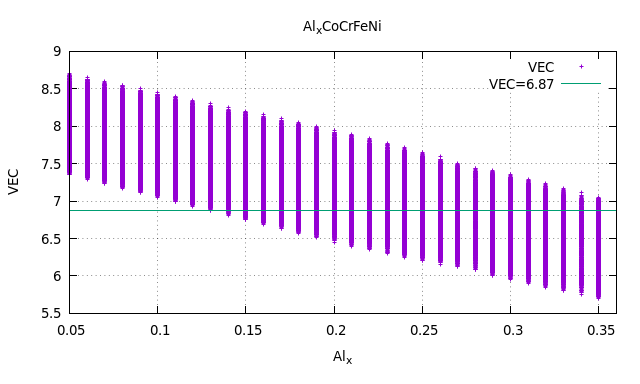} 
 
\begin{center} \small 
\emph{\footnotesize \textbf{Figura 11.} Resultado de la programación en FORTRAN90. Para la aleación $Al_xCoCrFeNi$ se graficó la concentración de electrones de valencia $VEC$ (ecuación 7) y el limite $VEC=6.87$ para la formación de estructuras de tipo BCC, lo que mostró una alta tendencia a formar estructuras de este tipo alrededor de las concentraciones $Al_{0.14}$ - $Al_{0.35}$. Liang y Schmid proponen que por debajo del limite $VEC<6.92$ se forman estructuras de dos fases $BCC + B2$. Este resultado es positivo de acuerdo con  Liang y Schmid \cite{liang2017}}.    
\end{center} 
\end{figure} 

\begin{figure}[!htb]   
\centering
\includegraphics[width=0.9\textwidth]{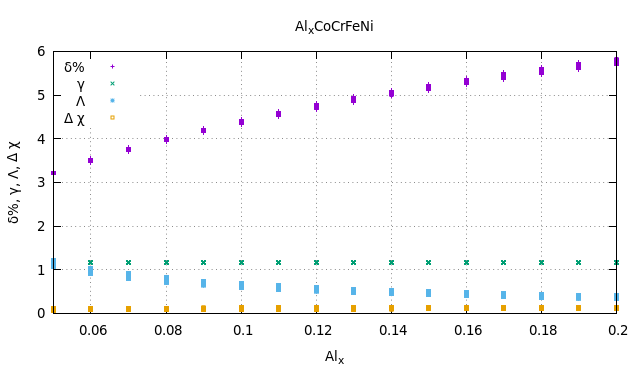} 
 
\begin{center} \small 
\emph{\footnotesize \textbf{Figura 12.}  Resultado de la programación en FORTRAN90. Se muestran los parámetros de la figura (10), para, $Al_x=0.05,...,0.2$}. 
\end{center} 
\end{figure}
\clearpage
De acuerdo con los resultados de Liang y Schmid, mostrados en la figura (13) \cite{liang2017} los resultados que se obtuvieron en este trabajo, mostrados en las figuras (10, 11 y 12) están en buen acuerdo. 

\begin{figure}[!htb]   
\centering
\includegraphics[width=0.6\textwidth]{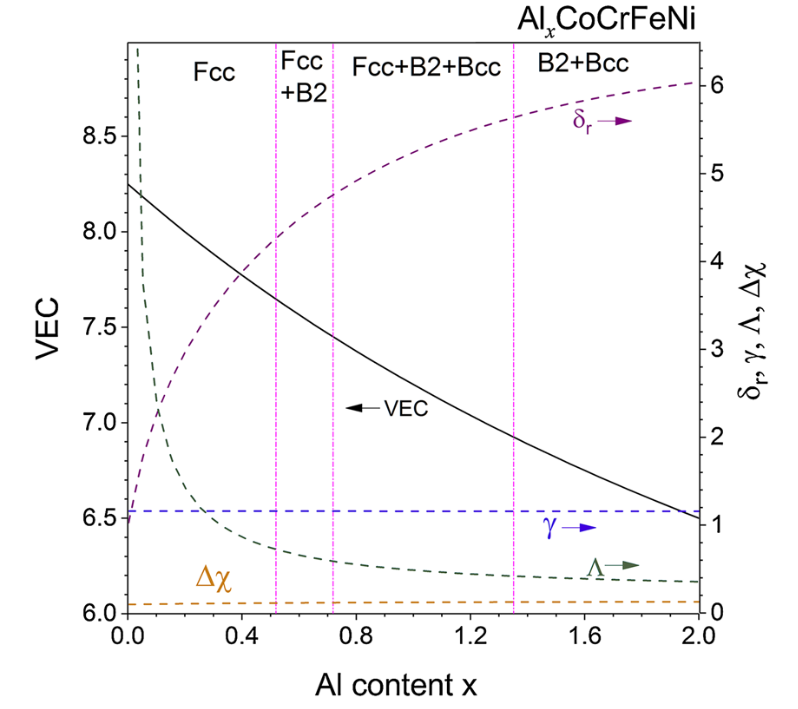} 
 
\begin{center} \small 
\emph{\footnotesize \textbf{Figura 13.}Resultados obtenidos por Liang y Schmid, aquí se muestran las ecuaciones (4,5,6,7 y 15) para la aleación $AlCoCrFeNi$, tomado de  \cite{liang2017}.}    
\end{center} 
\end{figure}  

Al comparar las figuras (10, 11, 12 y 13) notamos que existe un buen acuerdo entre los parámetros programados en FORTRAN90 $\delta\%, \gamma, \Lambda$ y $\Delta \chi$, y los resultados propuestos por Liang y Schmid \cite{liang2017}. A continuación observemos los resultados obtenidos con los parámetros $\Phi, \Omega, \Delta S_{mez}$ (figura 14) y $\Delta H_{mez}$ (figura 16).
\begin{figure}[!htb]   
\centering
\includegraphics[width=0.9\textwidth]{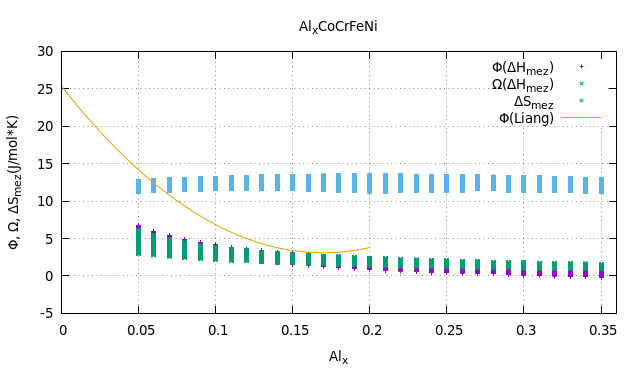} 

\begin{center} \small 
\emph{\footnotesize \textbf{Figura 14.} Para la aleación $AlCoCrFeNi$, se muestran los resultados obtenidos en FORTRAN90, aquí se muestran el parámetro $\Phi$   (ecuación 13), el parámetro $\Omega$ (ecuación 12), y la entropía configuracional $\Delta S_{mez}$ (ecuación 8), la linea continua de color naranja ($\Phi$) fue tomada punto a punto de la figura $15$, mediante el uso de una hoja de cálculo se encontró el siguiente ajuste polinomial, $\Phi(Liang)=773.058 Al^{2}_x - 262.515 Al_x + 25.377$, con coeficiente de determinación $R^2=0.958$.}
\end{center} 
\end{figure} 

\begin{figure}[!htb]   
\centering
\includegraphics[width=0.7\textwidth]{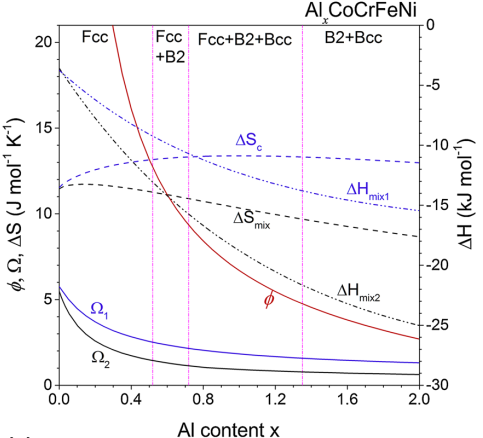} 
 
\begin{center} \small 
\emph{\footnotesize \textbf{Figura 15.}Resultados obtenidos por Liang y Schmid para la aleación $AlCoCrFeNi$, aquí se muestran en lineas azules, la entropía de mezcla ($\Delta S_c=\Delta S_{mez}$, ecuación 8), la entalpía de mezcla ($\Delta H_{mix1}=\Delta H_{mez}$, ecuación 9) el parámetro $\Omega$ ($\Omega_1=\Omega$, ecuación 12) y el parámetro $\Phi$  (linea roja, ecuación 13), las lineas negras hacen referencia a los mismos parámetros hallados por el método de $CALPHAD$, tomado de \cite{liang2017}.}    
\end{center} 
\end{figure}

\begin{figure}[!htb]   
\centering
\includegraphics[width=0.9\textwidth]{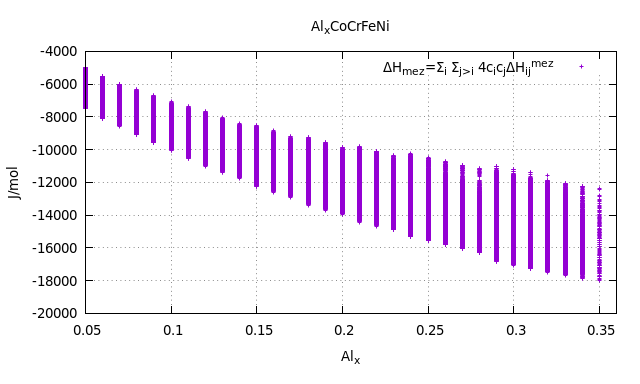} 
 
\begin{center} \small 
\emph{\footnotesize \textbf{Figura 16.} Para la aleación $Al_xCoCrFeNi$ se muestra la entalpía de formación $\Delta H_{mez}$ (ecuación 9) programada en FORTRAN90, comparar con la propuesta en la figura 15.}    
\end{center} 
\end{figure}

Comparando los valores de la entalpía de formación propuesta en la ecuación (9) (figuras 14, 15 y 16) la entropía configuracional dada en la ecuación (8) (figuras 14 y 15) el parámetro $\Omega$ dado en la ecuación (12) (figuras 14 y 15) evaluado con ayuda de la ecuación (9), se observó un buen acuerdo entre los valores propuesto por Liang y Schmid (lineas azules punteadas y continuas en la figura 15)  y los encontrados en este trabajo (figuras 14 y 16), pero aún no se acercan a los valores obtenidos por el método \emph{CALPHAD} hallados por este mismo grupo (lineas oscuras punteadas y continuas en la figura 15) quien es a la fecha la mejor aproximación a las medidas experimentales.   

A continuación se compararon los resultados experimentales propuesto por Wang \cite{wang2015} mostrados en la figura (17) y los encontrados con la ayuda de la programación (figura 18) para las aleaciones $Al_xCoCrFeNi$ y $TiVZrNbHf$, 

\begin{figure}[!htb]   
\centering
\includegraphics[width=0.7\textwidth]{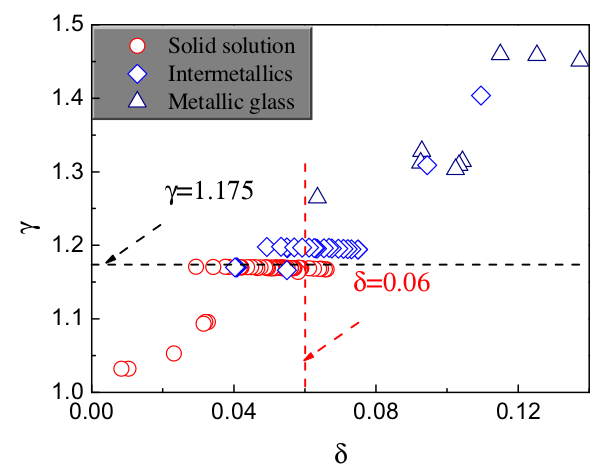} 
 
\begin{center} \small 
\emph{\footnotesize \textbf{Figura 17.} Comparación de los parámetros $\gamma$ y $\delta$. Se muestran un total de 95 aleaciones. El limite $\gamma=1.175$ distingue claramente las SSs de las fases intermetálicas y de los vidrios metálicos. Este limite arrojó un total de  59 SSs. Resultados experimentales tomados de Wang \cite{wang2015}}.    
\end{center}  
\end{figure}

\begin{figure}[!htb]   
\centering
\includegraphics[width=0.9\textwidth]{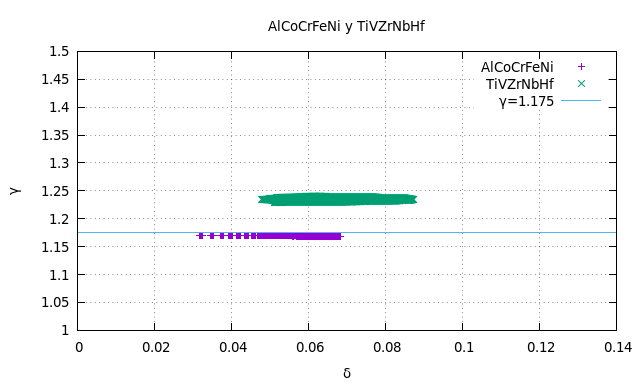} 
 
\begin{center} \small 
\emph{\footnotesize \textbf{Figura 18.} Para las aleaciones $AlCoCrFeNi$ y $TiVZrNbHf$, se muestran los parámetros $\gamma$ y $\delta$. El limite $\gamma=1.175$ mostró que la aleación $AlCoCrFeNi$ es candidata para formar HEAs.}     
\end{center}  
\end{figure}

De acuerdo con Wang \cite{wang2015} las SSs están mejor definidas por el parámetro $\gamma$ que por el parámetro $\delta$, esto se puede ver en la figura (17) cuyos resultados experimentales, mostraron que para un valor de $\delta$ entre $0.04$ y $0.06$, las SSs, los intermetálicos y lo vidrios metálicos pueden estar presente y no habría distinción entre alguna de estas fases, en cambio para un punto de $\gamma$, dichas fases estarían mejor definidas y muy poco se mezclan, pudiendo en el limite $\gamma=1.175$ separar las SSs de las otras fases.  

En el caso de los resultados hallados con ayuda de la programación en FORTRAN90 y el algoritmo de la figura (6), mostrados en la figura (18) se encontró una muy buena concordancia entre los valores de los parámetros $\gamma$ y $\delta$ para la aleación $AlCoCrFeNi$ y los hallados por Wang  mostrados en la figura (17). De acuerdo con nuestros resultados (figura 18) la aleación $TiVZrNbHf$ no mostró la formación de  SSs, encontrándose por encima del limite $\gamma=1.175$. Esto se puede entender si comparamos el radio atómico de mayor tamaño con los demás átomos que conforman cada aleación.  Para la aleación $AlCoCrFeNi$ el átomo de mayor radio atómico es el $Al$ con $143.17$ $pm$; sí comparamos este radio atómico con los demás que conforman la aleación, se encontró que, las diferencias entre radios atómicos no superan el $15$ $\%$, tal como se predicen las reglas de Hume-Rothery para las SSs binarias. Aquí la condición binaria es tomada de la comparación del radio atómico del $Al$ con los átomos restantes. 
Por otro lado la aleación $TiVZrNbHf$ al tener al $Zr$ como el elemento de mayor radio atómico  ($160.25$ $pm$), al compararlo con el  radio atómico del $V$ ($131.6$ $pm$) la diferencia es de $17.88$ $\%$ aproximadamente, violando la condición impuesta por Hume-Rothery para la diferencia máxima entre radios atómicos (15 $\%$) entre pares de especies. Esta es la causa por el cual las SSs están mejor definidas para la aleación $AlCoCrFeNi$ que para la aleación $TiVZrNbHf$ (ver figura $18$).

\subsection{Resultados que se deducen del exceso de entropía}

A pesar de que existe una buena similitud entre los resultados obtenidos con FORTRAN90 y los propuestos por Liang y Schmid para el $Al_xCoCrFeNi$ \cite{liang2017} excepto para el parámetro $\Phi$ (ver figura 14 y 15) se encontró que el exceso de entropía (ecuación 44) por ende las condiciones impuestas sobre la relación $S_{corr}/S_{id}$, son cumplidas \cite{he2016configurational, ding2018} los resultados de la programación se muestran en la figura (19) para el $Al_xCoCrFeNi$ y en la figura (20) para el $TiVZr_xNbHf$. Hay que aclarar que la definición del exceso de entropía $S_E$ que originalmente propuso Ye \cite{liang2017} junto con el parámetro $\Phi$,  no se usó en este trabajo, en cambio se usó la definición (ecuación 44) propuesta por He [Ding et al., 2018, He et al., 2016], siendo el motivo por el cual no se encontró en este trabajo un buen ajuste del parámetro $\Phi$, esto se hizo así porque en un principio la programación del exceso de entropía $S_E$ definido por Ye [Liang and Schmid-Fetzer, 2017] no dio los resultados esperados, arrojando valores nulos, por lo tanto se buscó una definición alternativa del $S_E$. Por su definición, se decidió escoger la definición ofrecida por He \cite{ding2018,he2016configurational} quien a nuestro juicio ofrece una definición correcta de $S_E$, gracias a que relaciona directamente la contribución química (Miedema) en las \emph{HEAs} cuando estas se han solidificado y están por debajo de su temperatura de fundición.

\begin{figure}[!htb]   
\centering
\includegraphics[width=0.9\textwidth]{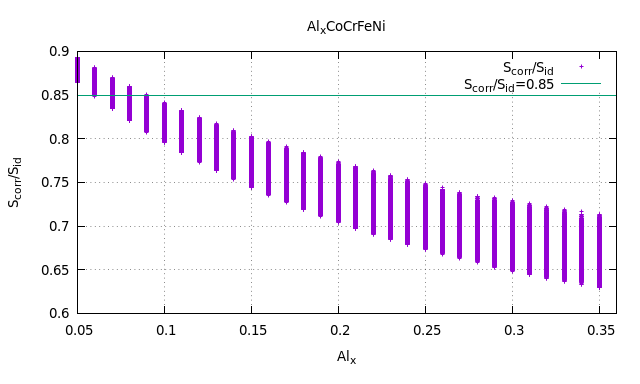} 
 
\begin{center} \small 
\emph{\footnotesize \textbf{Figura 19.} Para la aleación $Al_xCoCrFeNi$ se muestra la relación $S_{corr}/S_{id}$ y el  limite $S_{corr}/S_{id}=0.85$ (ver ecuación 43) programados en FORTRAN90 con ayuda del algoritmo de la figura (9). De acuerdo con el limite impuesto existen pocas \emph{HEAs}, por debajo de $Al_{0.09}$. El exceso de entropía se evaluó en la temperatura de fundición.}    
\end{center}  
\end{figure}

\begin{figure}[!htb]   
\centering
\includegraphics[width=0.9\textwidth]{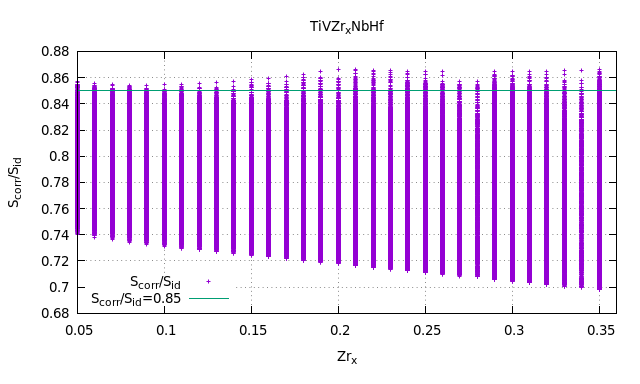} 
 
\begin{center} \small 
\emph{\footnotesize \textbf{Figura 20.} Para la aleación $TiVZr_xNbHf$ se muestra la relación $S_{corr}/S_{id}$ y el limite $S_{corr}/S_{id}=0.85$, programados en FORTRAN90. De acuerdo con el limite impuesto la posibilidad de encontrar \emph{HEAs} para la variación del $Zr_x$ son amplias. El exceso de entropía se evaluó en la temperatura de fundición.}    
\end{center} 
\end{figure}

Entonces para determinar el estado en que se encuentra la aleación puede referenciarse a los limites que ofrece el $VEC$ para la estructura cristalina, y a la razón $S_{corr}/S_{id}$ (ecuación 43), de acuerdo con He las \emph{HEAs} están definidas en el rango $0.85 < S_{corr}/S_{id}<1$, para múltiples estructura cristalinas esta razón se encuentran en el rango $0.7 < S_{corr}/S_{id}<0.85$, y aleaciones amorfas en el rango $0.4 < S_{corr}/S_{id}<0.6$ \cite{he2016configurational} mostrando para el $Al_xCoCrFeNi$ (figura 19) al parecer que, existe la posibilidad de que se formen \emph{HEAs}  por debajo de la concentración $Al_{0.09}$ (comparar con la figura 11 y la tabla 2, donde la propuesta es diferente y se acepta un mayor número de estructuras \emph{BCC}). Por otra parte en la aleación $TiVZr_xNbHf$ (figura 22) para la relación $S_{corr}/S_{id}$ se encontró una amplitud de \emph{HEAs} acorde con los resultados del $VEC$ mostrados en la figura (22) y la tabla (2). El exceso de entropía $S_E$ fue evaluado en la temperatura de fundición, esto mostró que ha dicha temperatura las interacciones interatómicas no son tan bajas y no deben ser ignoradas (ver figuras 19 y 20).

De acuerdo con el $VEC$  se mostró que en su mayoría las \emph{HEAs} que se deducen de la aleación $AlCoCrFeNi$ son de estructura cristalina combinada $(FCC + BCC)$ para el rango $6.87 \leq VEC < 8$ (ver tabla 2 y figura 11) aunque también es posible de acuerdo con la figura (11) encontrar estructuras separadas $BCC$ y $FCC$.    

Observemos como son los parámetros $\delta \%, \gamma, \Lambda, \Phi,S_{mez}$ y $\Delta H_{mez}$, en la aleación $TiVZr_xNbHf$,

\begin{figure}[!htb]   
\centering
\includegraphics[width=0.9\textwidth]{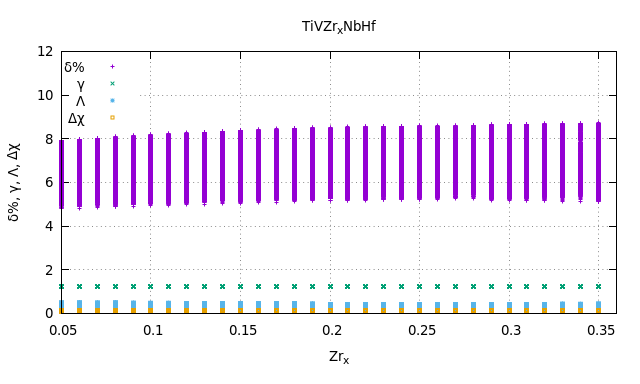} 

\begin{center} \small 
\emph{\footnotesize \textbf{Figura 21.} Parámetros $\delta\%$, $\gamma$, $\Lambda$, $\Delta\chi$, programados en FORTRAN90, para la aleación $TiVZr_xNbHf$. Se encontraron para esta aleación valores altos para el parámetro $\delta \%$, siendo este parámetro máximo ($\delta \%_{max}\approx 8.7\%$) para la \emph{HEA}, $Ti_{0.05}V_{0.35}Zr_{0.35}Nb_{0.13}Hf_{0.12}.$}  
\end{center} 
\end{figure}

\begin{figure}[!htb]   
\centering
\includegraphics[width=0.9\textwidth]{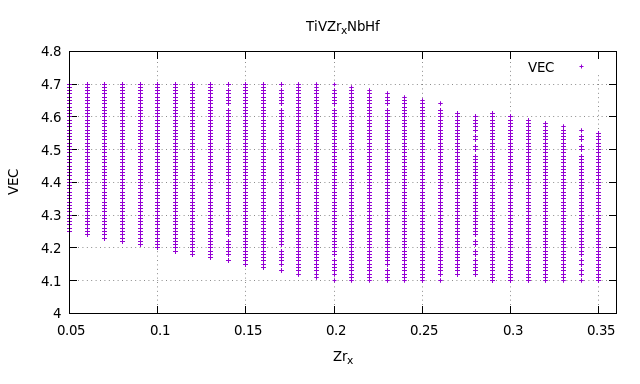} 

\begin{center} \small 
\emph{\footnotesize \textbf{Figura 22.}  Se muestra la concentración de electrones de valencia \emph{VEC} (ecuación 7) para la aleación $TiVZr_xNbHf$, la figura muestra que está aleación tiene una fuerte tendencia a formar estructuras \emph{BCC}. Ver las condiciones de la tabla 2.}  
\end{center} 
\end{figure}

\begin{figure}[!htb]   
\centering
\includegraphics[width=0.9\textwidth]{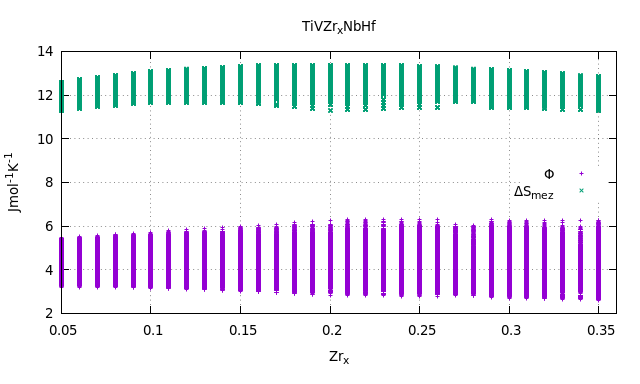} 

\begin{center} \small 
\emph{\footnotesize \textbf{Figura 23.}  Para la aleación $TiVZr_xNbHf$, se muestran los parámetros $\Phi$ y $\Delta S_{mez}$}. El parámetro $\Phi$ no cumple con la condición impuesta en la tabla (2), $\Phi>20$.  
\end{center} 
\end{figure}

\begin{figure}[!htb]   
\centering
\includegraphics[width=0.9\textwidth]{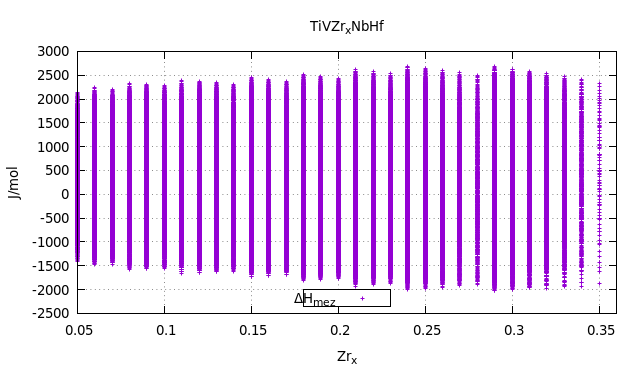} 
 
\begin{center} \small 
\emph{\footnotesize \textbf{Figura 24.}  Para la aleación $TiVZrNbHf$, se muestra la entalpía de formación $\Delta H_{mez}$ (ecuación 9),  se encontró una baja entalpía de formación, dentro de los limites establecidos en la formación de \emph{HEAs}, ver tabla 2. }
\end{center} 
\end{figure}

Los resultados de las figuras (21-24) mostraron un buen acuerdo para la aleación $TiVZrNbHf$ y los parámetros que se deducen de las reglas de Hume-Rothery y la termodinámica (excepto el parámetro $\Phi$), cabe destacar que se encontró un valor alto para el parámetro $\delta\% (\approx 8\%)$ (figura 17), además, de acuerdo con la figura 19, se predice que la mayoría de\emph{HEAs} de esta familia, formaran estructuras cristalinas tipo BCC, ideal para el almacenamiento de hidrógeno.

\subsection{Resultados que se deducen de la variación de la energía potencial elástica configuracional.}

De la ecuación (70) de acuerdo con Kamachali y Wang se debió encontrar que $\left.\frac{\Delta e \left (  c_i, c_j  \right )}{\Delta c_i}  \right |_j \rightarrow \sqrt{\lambda^2_{max}-\lambda^2_{min}} \approx 0.16$ \cite[figura 2a y 2b]{kamachali2022}, esta condición y la condición $\gamma=1.175$ \cite[figura 2]{wang2015} (ver tabla 2), sirven para diferenciar las SSs de las fases intermetálicas, amorfas y vidrios metálicos. 

A continuación se reescribió la ecuación (70), con $\Delta c_i=0$,

\begin{equation}
 \left.\frac{\Delta e \left (  c_i, c_j  \right )}{\Delta c_i}  \right |_j=q \left ( \lambda_{ih}^{2} - \lambda_{jh}^{2} \right )+2q \left (c_j \lambda_{jh}^{2} - c_i \lambda_{ih}^{2} - \lambda_{ih} \lambda_{jh} \left ( c_j - c_i \right ) \right ),  
 \end{equation} 
sí $\lambda_{ih}=\lambda_{jh}$ se sigue cumpliendo que $\left.\frac{\Delta e \left (  c_i, c_j  \right )}{\Delta c_i}  \right |_j=0 $. En la figura (20) se muestran cuales fueron los resultados obtenidos de la programación en FORTRAN90 y la ecuación (76) usando el algoritmo dado en la figura (6), haciendo, $\lambda_{ih}=\lambda_{max}=\lambda_{max}(Al)$, $\lambda_{jh}=\lambda_{min}=\lambda_{min}(Fe)$ para la aleación $AlCoCrFeNi$, y $\lambda_{ih}=\lambda_{max}=\lambda_{max}(Zr)$, $\lambda_{jh}=\lambda_{min}=\lambda_{min}(V)$ para la aleación $TiVZrNbHf.$

\begin{figure}[!htb]   
\centering
\includegraphics[width=0.9\textwidth]{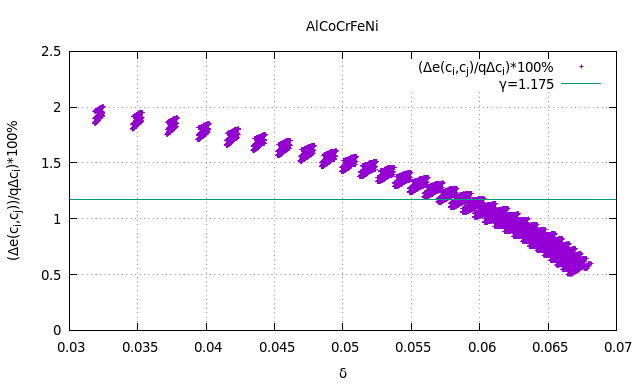} 
\begin{center} \small 
\emph{\footnotesize \textbf{Figura 25.}  Programación de la ecuación (76) y el parámetro $\delta$ en FORTRAN90 usando el algoritmo de la figura (9); como muestra la figura se hace necesario reescribir la ecuación (76) de la siguiente forma, $\left.\frac{\Delta e \left (  c_i, c_j  \right )}{q\Delta c_i}  \right |_j *100\%$. Se encontró un buen acuerdo entre la ecuación (76), y los limites $\gamma=1.175$ y $\delta\%\approx 6\%$, para la aleación $Al_{0.21}Co_{0.05}Cr_{0.05}Fe_{0.35}Ni_{0.34}$.  } 
\end{center} 
\end{figure}

Sí se reescribe la ecuación (76) como,

\begin{equation}
 \frac{1}{q}\left.\frac{\Delta e \left (  c_i, c_j  \right )}{\Delta c_i}  \right |_j= \left ( \left ( \lambda_{ih}^{2} - \lambda_{jh}^{2} \right )+2 \left (c_j \lambda_{jh}^{2} - c_i \lambda_{ih}^{2} - \lambda_{ih} \lambda_{jh} \left ( c_j - c_i \right ) \right ) \right )*100\%,  
 \end{equation}
se encontró que existe un buen acuerdo entre las condiciones $\gamma\leq1.175$  y   $\delta\% \approx 6\%$, a pesar de esto, de acuerdo con Kamachali y Wang la ecuación (76) debe ser equivalente a,

\begin{equation}
 \left.\frac{\Delta e \left (  c_i, c_j  \right )}{\Delta c_i}  \right |_j \approx q\sqrt{ \lambda_{max}^{2} - \lambda_{min}^{2}},  
 \end{equation}

 observemos los resultados obtenidos de haber programado la ecuación (78),

\begin{figure}[!htb]   
\centering
\includegraphics[width=0.9\textwidth]{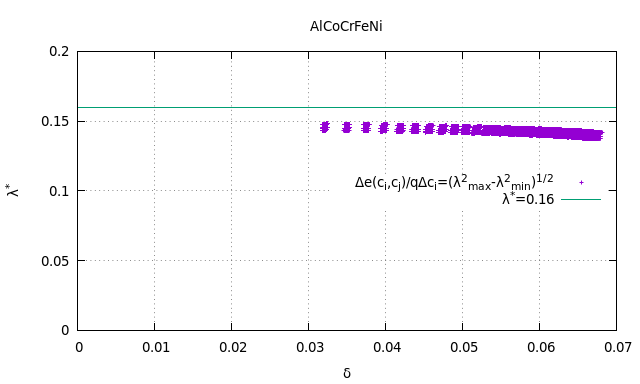} 
\begin{center} \small 
\emph{\footnotesize \textbf{Figura 26.}  Para la aleación $AlCoCrFeNi$ se muestra la programación de la ecuación (78) y el parámetro $\delta$ en FORTRAN90 donde se usó el algoritmo de la figura (9). También se muestra el limite propuesto Kamachali y Wang, $\lambda^*=0.16$, el cual separa las SSs de cualquier otra fase, como las intermetálicas y vidrios metálicos.} 
\end{center} 
\end{figure}

\begin{figure}[!htb]   
\centering
\includegraphics[width=0.9\textwidth]{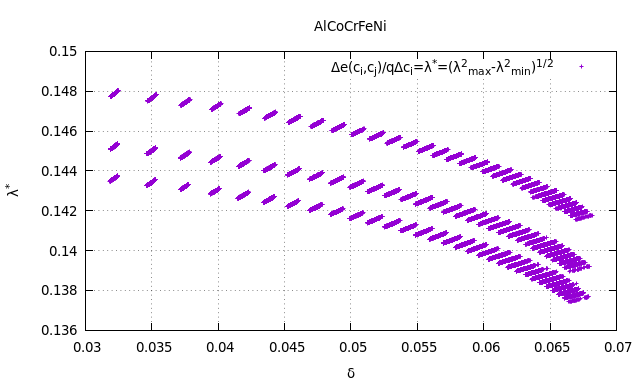} 
\begin{center} \small 
\emph{\footnotesize \textbf{Figura 27.} Para la aleación $AlCoCrFeNi$, se graficó la ecuación (78) y el parámetro $\delta $. Se muestra un acercamiento a la figura (26). La programación se hizo  en FORTRAN90 usando el algoritmo de la figura (9).}  
\end{center} 
\end{figure}

De los resultados mostrados por la aleación $AlCoCrFeNi$ (figuras 22-24)  se encontró un excelente acuerdo entre los resultados de Kamachali y Wang \cite{kamachali2022} y los simulados en este trabajo con la ayuda de FORTRAN90 y el algoritmo de la figura (9).   

Del comportamiento mostrado por la aleación $AlCoCrFeNi$ (figuras 25-27) hay que recalcar que el $Al$ es el elemento de mayor radio atómico ($\approx 143.17$ $pm$) y su diferencia con respecto a cada uno de los demás radios atómicos no supera el $15\%$, siendo este un buen acuerdo con la regla de Hume-Rothery que propone que para SSs binarias la diferencia entre radios atómicos no debe ser superior al $15\%$, en cambio, para la aleación $TiVZrNbHf$ teniendo al $Zr$ como elemento de mayor radio atómico ($\approx 160.25$ $pm$) su diferencia respecto al $V$ es $\approx 17.88$ $\%$ lo que haría que se formen compuestos; observemos como es el comportamiento de las ecuaciones (77 y 78) sobre la aleación $TiVZrNbHf$ (figuras 25-27).

\begin{figure}[!htb]   
\centering
\includegraphics[width=0.9\textwidth]{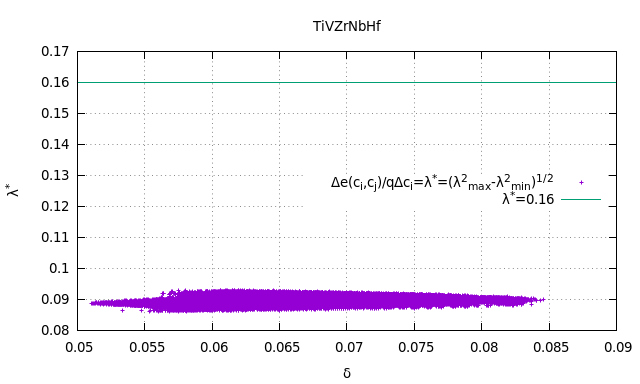} 
\begin{center} \small 
\emph{\footnotesize \textbf{Figura 28.} Para la aleación $TiVZrNbHf$, se graficó la ecuación (78), el parámetro $\delta $ y el limite $\lambda^*=0.16$. La programación se hizo  en FORTRAN90 usando el algoritmo de la figura (9).}  
\end{center} 
\end{figure}

\begin{figure}[!htb]   
\centering
\includegraphics[width=0.9\textwidth]{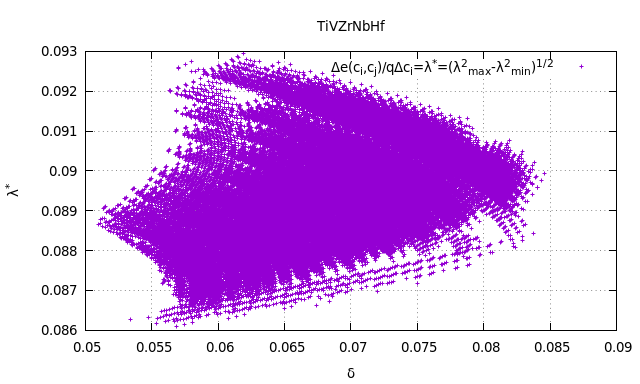} 
\begin{center} \small 
\emph{\footnotesize \textbf{Figura 29.} Para la aleación $TiVZrNbHf$, se graficó la ecuación (78) y el parámetro $\delta $. Se muestra un acercamiento a la figura (28). La programación se hizo  en FORTRAN90 usando el algoritmo de la figura (9).}  
\end{center} 
\end{figure}

\begin{figure}[!htb]   
\centering
\includegraphics[width=0.9\textwidth]{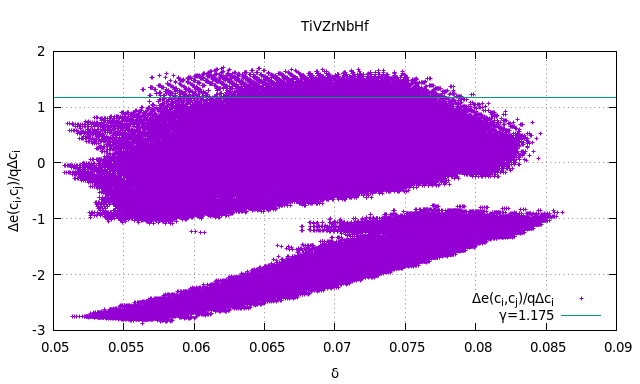} 
\begin{center} \small 
\emph{\footnotesize \textbf{Figura 30.} Para la aleación $TiVZrNbHf$, se graficó la ecuación (77), el parámetro $\delta$ y el limite $\gamma=1.175$. La programación se hizo  en FORTRAN90 usando el algoritmo de la figura (9).}  
\end{center} 
\end{figure}

Para la aleación $TiVZrNbHf$ la figura (28) muestra el comportamiento de la ecuación (78), de acuerdo con esto se siguió respetando el limite $\lambda^* \leq 0.16$ \cite{kamachali2022} para separar las SSs de cualquier otra fase. La ventaja de la programación fue que permitió al igual como en la figura (27), hacer un acercamiento mostrado en la figura (28), esto se muestra en la figura (26). La figura (30) muestra la programación de la ecuación (77) que es el resultado completo de nuestro trabajo, al igual que la figura (25), la figura (30) mostró puntos que están en el rango $\gamma \leq 1.175$ haciendo que la variación de la energía potencial elástica configuracional dada en la ecuación (77) y el parámetro $\gamma$ puedan ser usados como criterio para fabricar \emph{HEAs}.

\subsection{Resultados que se deducen del modelo para la entalpía de mezcla y corregida.}

De la programación en FORTRAN90 y el modelo teórico para la entalpía de formación $\Delta H_{mez}^*$ y corregida $\Delta H_{corr}$, ecuaciones (68) y (69) respectivamente, se encontró al comparar con la figura 15, que solo el término $\sum_{k=1}^{N}s_kV_ke$ y la corrección química (ver el termino con raíz de la ecuación 66) son las que contribuye a la entalpía de formación, por lo tanto en un primer acercamiento se propuso que, 

\begin{equation}
    \Delta H_{mez}^*= \begin{cases}
 & \sum_{k=1}^{N}s_kV_ke,  \text{ sí } \bar{H} \geq 0 \\
 &   -\sum_{k=1}^{N}s_kV_ke, \text{ sí } \bar{H} < 0,
\end{cases}
\end{equation}
obteniendo el resultado que se muestra en la figura 31.

\begin{figure}[!htb]   
\centering
\includegraphics[width=0.9\textwidth]{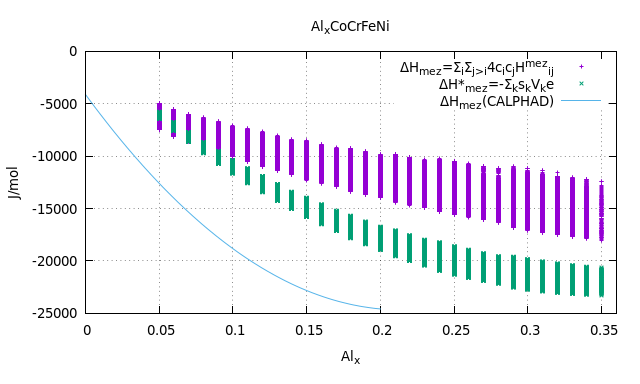} 
\begin{center} \small 
\emph{\footnotesize \textbf{ Figura 31.} Para la aleación $Al_xCoCrFeNi$ se comparan las entalpías teórica ($\Delta H_{mez}=\sum_{i}^{N-1}\sum_{j>i}^{N}4c_ic_jH_{ij}^{mez}$, ecuación 9), teórica-experimental (método de CALPHAD, ver figura 15 de la cual se obtuvo con la ayuda de una hoja de calculo el siguiente ajuste polinomial, $\Delta H_{mez}(CALPHAD)=448463.199 Al_{x}^2-192042.649 Al_x -4109.200$, con coeficiente de determinación $R^2=0.999$, con bases a los puntos tomados a mano de la figura 15, para $\Delta H_{mix2}$) y la  propuesta de este trabajo (ecuación 79) con, $\Delta H_{mez}^*=-\sum_{k=1}^{N}s_kV_ke$, siendo $s_k=\frac{2}{3}\cdot \frac{1}{3-\frac{1}{\frac{1-\nu_k}{1+\nu_k}}}$, donde $\nu_k$ es el cociente de Poisson, reescrita de forma análoga a la ecuación (50).}  
\end{center} 
\end{figure}

\begin{figure}[!htb]   
\centering
\includegraphics[width=0.9\textwidth]{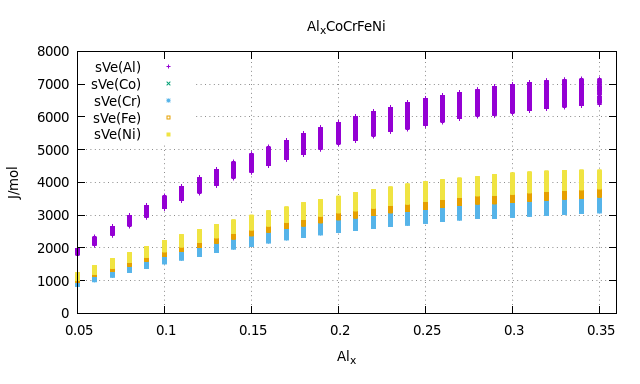} 
\begin{center} \small 
\emph{\footnotesize \textbf{Figura 32.} Para la aleación $Al_xCoCrFeNi$ se muestra la contribución de cada especie atómica a la energía potencial elástica configuracional (ecuación 80). Se observó una fuerte contribución a dicha energía por parte del aluminio, esto es así debido a que el aluminio es el elemento químico de mayor radio atómico, siendo $143.17$ $pm$, $125.10$ $pm,$  $124.91$ $pm$, $124.12$ $pm$ y  $124.59$ $pm$, los radios atómicos del $Al, Co, Cr, Fe y Ni$, respectivamente  \cite{miracle2017}. }  
\end{center} 
\end{figure}

\newpage

\begin{figure}[!htb]   
\centering
\includegraphics[width=0.9\textwidth]{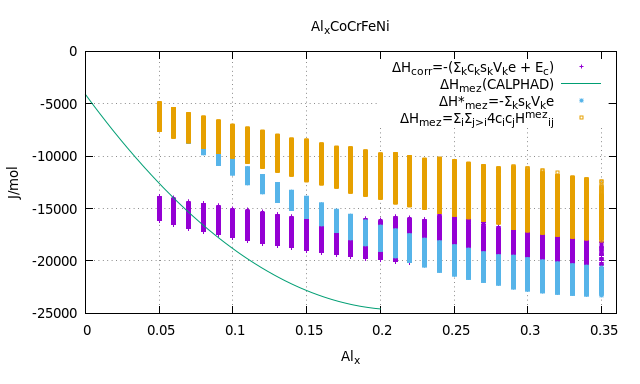} 
\begin{center} \small 
\emph{\footnotesize \textbf{Figura 33.} Entalpía de formación para el $Al_xCoCrFeNi$ con $\Delta H_{corr}=-\left (\sum_{k=1}^{N}c_ks_kV_ke +E_c \right )$, siendo $s_k=\frac{2}{3}\cdot \frac{1}{3-\frac{1}{\frac{1-\nu_k}{1+\nu_k}}}$, reescrita de forma análoga a la ecuación (50), donde $\nu_k$ es el cociente de Poisson de la sustancia $k$, $E_c=\sqrt{\sum_{i}^{N-1}\sum_{j>i}^{N}c_ic_j(H_{ij}-\Bar{H})^2}$. Este resultado está en buen acuerdo con los propuestos en la tabla (2) y los hallados por el método de \emph{CALPHAD} (ver figura 15, para $\Delta H_{mix2}$).}  
\end{center} 
\end{figure}

De la figura (31 y 33) se observó un buen acercamiento al limite propuesto por Yang para la entalpía de formación, $-15 \hspace{1mm} kJ/mol \leq \Delta H_{mez} \leq 5 \hspace{1mm} kJ/mol$ [Yang and Zhang, 2012, Liang and Schmid-Fetzer, 2017]  y al propuesto por el método de \emph{CALPHAD} $-25 \hspace{1mm} kJ/mol \leq \Delta H_{mez} \leq -5 \hspace{1mm} kJ/mol$ (para la aleación $AlCoCrFeNi$, ver figura 15) \cite{liang2017}. Aquí se hizo necesario la introducción de $s_k$, definida como $s_k=\frac{2}{3}\cdot \frac{1}{3-\frac{1}{\frac{1-\nu_k}{1+\nu_k}}}$, ver ecuación (50). 
     
En la figura (31) se compararon tres formas diferentes de la entalpía, siendo la ecuación (79) nuestro aporte con el trabajo, la entalpía de mezcla (Miedema) la ecuación (9) y la teórico experimental (figura 15) $\Delta H (CALPHAD)$ que se deduce del modelo CALPHAD \cite[ver figura 1c]{liang2017}.

La figura (32) estableció la contribución de la energía potencial elástica configuracional de cada especie atómica a la entalpía de formación (ecuación 79), aquí se encontró que el aluminio al ser el elemento químico de mayor radio atómico (en la aleación $AlCoCrFeNi$ \cite{miracle2017})  es quien más contribuye a dicha energía.

Se introdujo la corrección química (figura 33), que se da cuando la  \emph{HEA} (o aleación) se ha fabricado y está por debajo de su temperatura de fundición, entonces es cuando es posible que las interacciones interatómicas puedan formar compuestos químicos \cite{ding2018}, esta corrección se midió por los datos para la entalpía de formación binaria a partir del modelo de Miedema \cite{takeuchi2005classification}, se propuso que la entalpía corregida estuviera definida como,

\begin{equation}
 \Delta H_{corr}= \begin{cases}
 & \sum_{k=1}^{N}s_kV_ke,  \text{ sí } \bar{H} \geq 0 \\
 &   -\left (  \sum_{k=1}^{N}c_ks_kV_ke + \sqrt{\sum_{i}^{N-1}\sum_{j>i}^{N}c_ic_j(H_{ij}-\Bar{H})^2} \right ), \text{ sí } \bar{H} < 0,
\end{cases}
\end{equation}
donde, aparte de introducir dicha corrección química, también se colocaron las proporciones molares $c_k$, arrojando el resultado de la figura (33), estando este resultado dentro de los limites aceptados, ver la entalpía de formación de la tabla (2) y la figura (15). $\bar{H}$ es el promedio de los valores $H_{ij}$. 

El signo positivo de la ecuación (80) en un principio era negativo pero no se hallaron resultados validos, así que el signo positivo se debe a que si $\bar{H}<0$ entonces no solo tenemos la contribución de la ecuación (79), también hay que sumarle la contribución química cuando el proceso es exotérmico. El signo positivo también puede ser justificado si consideramos que la aleación o \emph{HEA} ya se ha fabricado y se encuentra por debajo de su temperatura de fundición (ecuación 10), una vez logrado esto si quisiéramos separar todas las componentes de la aleación no solo debemos agregar la entalpía de la ecuación (79) si no que también debemos sumar la energía necesaria para romper los enlaces químicos tal como muestra la ecuación (80) cuando  $\bar{H}<0$, dicha corrección no se tiene en cuenta si el proceso es endotérmico ($\bar{H}>0$) o sea, si la energía solo alcanza para ubicarlas en sus posiciones y no para enlazarlos químicamente. 

La entalpía de formación y el parámetro $\Omega$ se relacionan por medio de la expresión dada en la ecuación (12),  en las figuras (34 y 35) se muestran el parámetro $\Omega$, comparado para tres entalpías de formación diferentes (ecuación 9, 79 y 80),  para las aleaciones $TiVZr_xNbHf$ y $Al_xCoCrFeNi$ respectivamente. En nuestro trabajo el parámetro $\Omega$ se evaluó a la temperatura de fundición.

\begin{figure}[!htb]   
\centering
\includegraphics[width=0.9\textwidth]{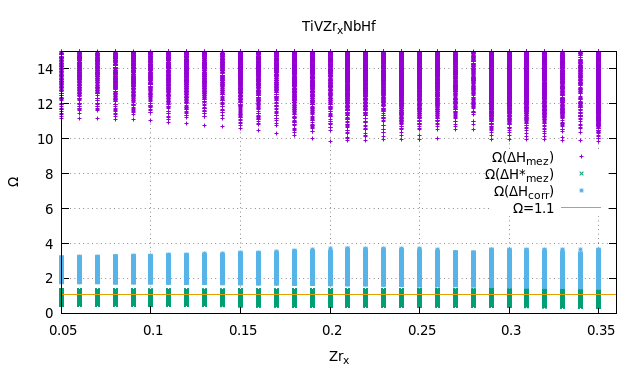}
\begin{center} \small 
\emph{\footnotesize \textbf{Figura 34.}  Para la aleación $TiVZr_xNbHf$, se muestra el parámetro $\Omega$ (ecuación 12),  evaluado para tres entalpías diferentes $\Delta H_{mez}$ (ecuación 9), $\Delta H*_{mez}$ (ecuación 79) y $\Delta H_{corr}$ (ecuación 80).}
\end{center} 
\end{figure}

\newpage

\begin{figure}[!htb]   
\centering
\includegraphics[width=0.9\textwidth]{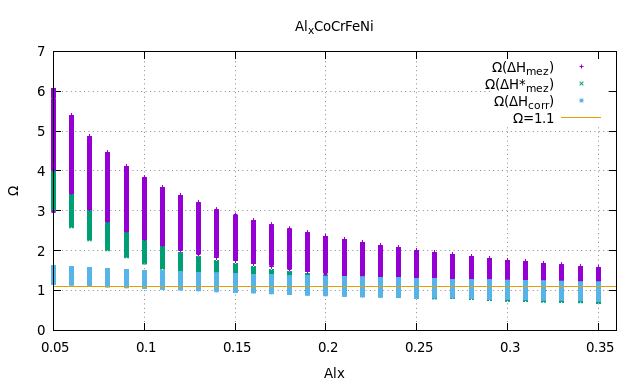}
\begin{center} \small 
\emph{\footnotesize \textbf{Figura 35.}  Para la aleación $Al_xCoCrFeNi$, se muestra el parámetro $\Omega$ (ecuación 12),  evaluado para tres entalpías diferentes $\Delta H_{mez}$ (ecuación 9), $\Delta H*_{mez}$ (ecuación 79) y $\Delta H_{corr}$ (ecuación 80)}.
\end{center} 
\end{figure}

De acuerdo con Liang y Schmid\cite{liang2017} las \emph{HEAs} existen para el rango $\Omega>1.1$, como podemos ver en las figuras (34 y 35) este limite se respeta independientemente de la definición que usemos para la entalpía de formación (ecuaciones 9, 79 y 80), la figura (34) mostró que, el parámetro $\Omega$ para la aleación $TiVZr_xNbHf$ se acerca al limite $\Omega=1.1$, cuando la entalpía de formación estuvo definida por la ecuación (79), tomando como contribución solamente la energía elástica configuracional. Por otro lado la figura (35) mostró que la aleación $Al_xCoCrFeNi$ se acerca al limite  $\Omega=1.1$ cuando la entalpía de formación estuvo definida por la ecuación (80),  donde se toman como contribuciones la energía potencial elástica  configuracional y la corrección química hecha cuando los elementos forman compuestos. Se debe a que según la figura (24) la aleación $TiVZr_xNbHf$ posee una baja entalpía de formación (en algunos casos el proceso es endotérmico) esto indica que la corrección química dada en la ecuación (80) no tiene relevancia y de acuerdo con $\Omega$ la entalpía de formación dada por la ecuación (79) es un buen acercamiento. En cambio la aleación $Al_xCoCrFeNi$ como indicó la figura (33) poseyó una mayor entalpía de formación (en su mayoría el proceso es exotérmico) esto revela que el parámetro $\Omega$ debió estar mejor definido por la ecuación (80) tal como mostró la figura (35).

\newpage

\section{Conclusiones}
Para las aleaciones $AlCoCrFeNi$ y $TiVZrNbHf$, se encontró un buen acuerdo entre los parámetros que se deducen de las reglas de Hume Rothery, la termodinámica (tabla 2). En la aleación $AlCoCrFeNi$ se encontró que la diferencia entre radios atómicos, comparados con el $Al$ no superó el $15$ $\%$ propuesto por Hume-Rothery, esto tiene como consecuencia que el limite $\gamma=1.175$ es muy bueno para distinguir las SSs que esta aleación pueda a llegar a formar, esto se muestra en las figuras (18, 25 y 26). La variación de la energía potencial elástica configuracional  mostró, que, es correcta la hipótesis de que existe un módulo de comprensibilidad equivalente a toda las especies atómicas, permitiendo definir la entalpía de formación (ecuaciones 79 y 80) aproximadamente dentro de los limites establecidos por el método de \emph{CALPHAD} $ -5$ $kJ/mol$ $\leq \Delta H_{mez} \leq  -25$ $kJ/mol$ (ver figura 15) para la aleación $Al_xCoCrFeNi$. Debido al \emph{VEC} y a la condición $0.85 \leq S_{corr}/S_{id} \ 1$, se propuso que la aleación $TiVZrNbHf$ deberá presentar más estructuras $BCC$ que las que puede presentar la aleación $AlCoCrFeNi$, esto hace candidata a la aleación $TiVZrNbHf$ para futuras aplicaciones de almacenamiento de hidrógeno, la aleación $AlCoCrFeNi$ también mostró estructuras tipo $BCC$, pero lo hizo en menor cantidad (ver figuras 11, 22, 19 y 20); según Liang y Schmid \cite{liang2017} encontraron de acuerdo con el método de CALPHAD que las estructuras reales serian $BCC + B2$. El exceso de entropía $S_E$ que aparece en las ecuaciones (43 y 44) se tomó para la temperatura de fundición, mostrando que a pesar de la fuerte agitación térmica, en estado físico los átomos si interactúan, formando compuestos. Contrario de lo que se pensaba, debido a que la teoría predice que el fuerte agitamiento producido por la temperatura de fundición no permitiría la interacción interatómica.

\newpage

\footnotesize{}

\bibliographystyle{apalike}
\bibliography{Royero}

\newpage 
 \section{Apéndice A} 

En la ecuación (40) encontramos la descripción de la energía potencial elástica configuracional considerada elásticamente isótropas, 

 \begin{equation} \tag{A1}
 e = q   \sum_{\alpha\neq h}^{N} \lambda_{\alpha h}^2 \left ( c_\alpha- c_\alpha^2 \right )- 2q   \sum_{\scriptsize \begin{matrix}
 \alpha,\beta \neq h \\       
  \alpha\neq \beta 
\end{matrix}}^{N} \lambda_{\alpha h} \lambda_{\beta h} c_{\alpha} c_{\beta},   
\end{equation}
donde $q=2\Bar{\mu} \left ( \frac{1+\Bar{\nu}}{1-\Bar{\Bar{\nu}}}\right )$.

A continuación reescribimos la ecuación (5),
\begin{equation}\tag{A2}
    \delta^2=\sum_{\alpha = i}^{N}c_\alpha \left ( \frac{r_\alpha - \bar r}{\bar r}  \right )^2=\sum_{\alpha = i}^{N}c_\alpha \left ( \frac{r_\alpha - r_h}{\bar r} + \frac{r_h - \bar r}{\bar r}  \right )^2=\sum_{\alpha = i}^{N}c_\alpha \left ( \lambda_{\alpha h} + \frac{r_h - \bar r}{\bar r}  \right )^2,
\end{equation}
 donde se usó, $\lambda_{\alpha h}=\frac{r_\alpha - r_h}{\bar r}$, continuemos desarrollando la ecuación (A2),
 
\begin{equation}\tag{A3}
    \delta^2=\sum_{\alpha \neq h}^{N}c_\alpha \lambda_{\alpha h}^2 + 2  \left ( \frac{r_h - \bar r}{\bar r} \right )\sum_{\alpha \neq h}^{N}c_\alpha \lambda_{\alpha h} + \left ( \frac{r_h - \bar r}{\bar r} \right )^2,
\end{equation}
 aquí se tuvo en cuenta la condición, $\sum_{\alpha=i}^{N}c_\alpha=1$, si hacemos $  \frac{r_h - \bar r}{\bar r}=-\sum_{\alpha \neq h}^{N}c_\alpha \lambda_{\alpha h}$ \cite{kamachali2022}, se encontró la siguiente definición,

\begin{equation}\tag{A4}
    \delta^2=\sum_{\alpha \neq h}^{N}c_\alpha \lambda_{\alpha h}^2 - \left ( \frac{r_h - \bar r}{\bar r} \right )^2,
\end{equation} 
desarrollemos el segundo miembro de la ecuación A4,

\begin{equation}\tag{A6}
    \left ( \frac{r_h - \bar r}{\bar r} \right )^2=\left (  - \sum_{\alpha \neq h}^{N}c_\alpha \lambda_{\alpha h}\right ) \left (  - \sum_{\beta \neq h}^{N}c_\beta \lambda_{\beta h}\right )
\end{equation} 

\begin{equation}\tag{A7}
    =\left ( c_i\lambda_{ih} + c_j\lambda_{jh} + ... + c_N \lambda_{Nh} \right ) \left ( c_i\lambda_{ih} + c_j\lambda_{jh} + ... + c_N \lambda_{Nh} \right )
\end{equation} 

\begin{equation}\tag{A8}
    =\left ( \left ( c_i\lambda_{ih} \right )^2+ \left ( c_j\lambda_{jh}\right )^2 + ... + \left ( c_N\lambda_{Nh}\right )^2 + 2c_i c_j \lambda_{ih}\lambda_{jh} + ... + 2c_{N-1} c_N \lambda_{(N-1) h}\lambda_{N h} \right )
\end{equation} 

\begin{equation}\tag{A9}
 \left ( \frac{r_h - \bar r}{\bar r} \right )^2  =   \sum_{\alpha \neq h}^{N} \left ( c_\alpha \lambda_{\alpha h}\right )^2 + 2\sum_{\scriptsize \begin{matrix}
 \alpha,\beta \neq h \\       
  \alpha\neq \beta 
\end{matrix}}^{N} c_\alpha c_\beta \lambda_{ \alpha h} \lambda_{\beta h},
\end{equation} 
 reemplazando la ecuación (A9) en la ecuación (A4) se obtiene, 

\begin{equation}\tag{A10}
    \delta^2=\sum_{\alpha \neq h}^{N}c_\alpha \lambda_{\alpha h}^2 -  \sum_{\alpha \neq h}^{N} \left ( c_\alpha \lambda_{\alpha h}\right )^2 - 2\sum_{\scriptsize \begin{matrix}
 \alpha,\beta \neq h \\       
  \alpha\neq \beta 
\end{matrix}}^{N} c_\alpha c_\beta \lambda_{ \alpha h} \lambda_{\beta h}.
\end{equation}

 Sí comparamos las ecuaciones (A1) y (A10) llegamos finalmente a la siguiente definición,
\begin{equation}\tag{A11}
   e=q\delta^2, 
\end{equation} 
 con lo cual se obtuvo la definición de energía potencial elástica configuracional para las \emph{HEAs} consideradas elásticamente isótropas (ver ecuaciones 40 y 41).

\section{Apéndice B}
Teniendo en cuenta que la variación en la concentración $\Delta c_\alpha$ produce una variación en el potencial $e$, de la siguiente forma, 

\begin{equation}\tag{B1}
 \Delta  e \left (  c_\alpha, c_\beta  \right ) =   e \left (  c_\alpha + \Delta c_\alpha , c_\beta + \Delta c_\beta  \right ) - e \left (  c_\alpha, c_\beta \right )
 \end{equation}

\begin{equation}\tag{B2}
 \Delta  e \left (  c_\alpha, c_\beta  \right ) =   e \left (  c_\alpha+ \Delta c_\alpha , c_\beta - \Delta c_\alpha  \right ) - e \left (  c_\alpha, c_\beta \right ),
 \end{equation} 
 dicha variación obedece la sustitución binaria $\alpha \rightarrow \beta$ esquematizada en la figura 3 (como $i \rightarrow j$), con $\Delta c_\alpha = - \Delta c_\beta $. De acuerdo con Kamachali las \emph{HEAs} consideradas isotópicamente elástica (estos es, se considera que existe un módulo de compresibilidad equivalente para todas las sustancias) están bien definidas por \cite{kamachali2022},
 
  \begin{equation}\tag{B3} 
 e = q   \sum_{\alpha\neq h}^{N} \lambda_{\alpha h}^2 \left ( c_\alpha- c_\alpha^2 \right )- 2q   \sum_{\scriptsize \begin{matrix}
 \alpha,\beta \neq h \\       
  \alpha\neq \beta 
\end{matrix}}^{N} \lambda_{\alpha h} \lambda_{\beta h} c_{\alpha} c_{\beta},   
\end{equation}
 
 A continuación sustituimos la ecuación (B3) en la ecuación (B2), 
 
 \begin{equation*}
 \Delta e \left (  c_\alpha, c_\beta  \right ) = q \sum_{\alpha \neq h}^{N} \lambda_{\alpha h}^{2} \left (c_\alpha + \Delta c_\alpha - \left ( c_\alpha + \Delta c_\alpha \right )^2 \right ) -2q \sum_{\scriptsize \begin{matrix}
 \alpha,\beta \neq h \\    
  \alpha \neq \beta 
\end{matrix}}^{N} \lambda_{\alpha h} \lambda_{\beta h} \left (c_\alpha + \Delta c_\alpha \right ) \left (c_\beta - \Delta c_\alpha \right ) 
\end{equation*} 
 
\begin{equation}\tag{B4}
 \hspace{-3.2cm}- q \sum_{\alpha \neq h}^{N} \lambda_{ih}^{2} \left (c_\alpha -c_\alpha^2 \right ) +2q \sum_{\scriptsize \begin{matrix}
 \alpha,\beta \neq h \\     
  \alpha \neq \beta
\end{matrix}}^{N} \lambda_{\alpha h} \lambda_{\beta h}c_\alpha c_\beta,  
\end{equation} 
reorganizando, 
 
 \begin{equation*}
 \Delta e \left (  c_\alpha, c_\beta  \right ) = q \sum_{\alpha \neq h}^{N} \lambda_{\alpha h}^{2} \left ( c_\alpha-c_\alpha^2 + \Delta c_\alpha -  2c_\alpha \Delta c_\alpha - \left (\Delta c_\alpha \right )^2 \right ) 
 \end{equation*}
 \begin{equation*}
 \hspace{3.3cm} -2q  \sum_{\scriptsize \begin{matrix}
 \alpha,\beta \neq h \\    
  \alpha \neq \beta  
\end{matrix}}^{N} \lambda_{\alpha h} \lambda_{\beta h} \left (c_\alpha c_\beta - c_\alpha\Delta c_\alpha + c_\beta \Delta c_\alpha - \left (\Delta c_\alpha \right )^2  \right ) 
\end{equation*} 
\begin{equation}\tag{B5}
 \hspace{2cm}  - q \sum_{\alpha \neq h}^{N} \lambda_{\alpha h}^{2} \left (c_\alpha -c_\alpha^2 \right ) +2q \sum_{\scriptsize \begin{matrix}
 \alpha,\beta  \neq h \\     
  \alpha \neq \beta
\end{matrix}}^{N} \lambda_{\alpha h} \lambda_{\beta h}c_\alpha c_\beta.  
\end{equation} 
  
A continuación reescribimos la ecuación (B5) usando la siguiente propiedad de linealidad de las sumatorias, $ \sum_{\alpha}^{N} \left ( a_\alpha + b_\alpha \right )= \sum_{\alpha}^{N} a_\alpha + \sum_{\alpha}^{N} b_\alpha  $, se tiene que, 

 \begin{equation*} 
 \Delta e \left (  c_\alpha, c_\beta  \right ) = q \sum_{\alpha \neq h}^{N} \lambda_{\alpha h}^{2} \left ( c_\alpha -c_\alpha ^2  \right ) +  q \sum_{\alpha \neq h}^{N} \lambda_{\alpha h}^{2} \left (\Delta c_\alpha -  2c_\alpha \Delta c_\alpha - \left (\Delta c_\alpha \right )^2 \right ) -2q  \sum_{\scriptsize \begin{matrix}
 \alpha,\beta \neq h \\     
  \alpha \neq \beta  
\end{matrix}}^{N} \lambda_{\alpha h} \lambda_{\beta h}  c_\alpha c_\beta
 \end{equation*}
 
 \begin{equation}\tag{B6}
 \hspace{-0.1cm}   - 2q  \sum_{\scriptsize \begin{matrix}
 \alpha,\beta \neq h \\   
  \alpha \neq \beta    
\end{matrix}}^{N} \lambda_{\alpha h} \lambda_{\beta h} \left ( -c_\alpha\Delta c_\alpha + c_\beta \Delta c_\alpha - \left (\Delta c_\alpha \right )^2  \right ) - q \sum_{\alpha \neq h}^{N} \lambda_{\alpha h}^{2} \left (c_\alpha -c_\alpha^2 \right ) +2q \sum_{\scriptsize \begin{matrix}
 \alpha,\beta \neq h \\     
  \alpha \neq \beta
\end{matrix}}^{N} \lambda_{\alpha h} \lambda_{\beta h}c_\alpha c_\beta.
\end{equation} 

La ecuación (B6) se simplifica a, 

 \begin{equation*} 
 \Delta e \left (  c_\alpha, c_\beta  \right ) =  q \sum_{\alpha \neq h}^{N} \lambda_{\alpha h}^{2} \left (\Delta c_\alpha -  2c_\alpha \Delta c_\alpha - \left (\Delta c_\alpha \right )^2 \right ) 
 \end{equation*} 
 \begin{equation}\tag{B7}
 \hspace{4cm} - 2q  \sum_{\scriptsize \begin{matrix}
 \alpha,\beta \neq h \\    
  \alpha \neq \beta   
\end{matrix}}^{N} \lambda_{\alpha h} \lambda_{\beta h} \left ( -c_\alpha \Delta c_\alpha + c_\beta \Delta c_\alpha - \left (\Delta c_\alpha \right )^2  \right ) 
\end{equation} 
   
Expandamos la primera sumatoria del lado derecho de la ecuación (B7) hasta $i$ y $j$, y la segunda sumatoria hasta $i, j$, como sigue,       
 
 \begin{equation*} \hspace{-2cm}
 \Delta e \left (  c_i, c_j  \right ) = q \lambda_{ih}^{2} \left (\Delta c_i -  2c_i\Delta c_i - \left (\Delta c_i \right )^2 \right ) +  q \lambda_{jh}^{2} \left (\Delta c_j -  2c_j\Delta c_j - \left (\Delta c_j \right )^2 \right )
 \end{equation*} 
 \begin{equation*}       
 \hspace{-1.3cm} + q \sum_{\alpha \neq i,j,h}^{N} \lambda_{\alpha h}^{2} \left (\Delta c_\alpha -  2c_\alpha \Delta c_\alpha - \left (\Delta c_\alpha \right )^2  \right ) - 2q  \lambda_{ih} \lambda_{jh} \left ( -c_i\Delta c_i + c_j \Delta c_i - \left (\Delta c_i \right )^2  \right )
 \end{equation*}
 
 \begin{equation}\tag{B8}
 \hspace{-5.4cm} - 2q \sum_{\scriptsize \begin{matrix} 
\alpha,\beta \neq i,j,h \\    
  \alpha \neq \beta   
\end{matrix}}^{N} \lambda_{\alpha h} \lambda_{\beta h} \left ( -c_\alpha\Delta c_\alpha + c_\beta \Delta c_\alpha - \left (\Delta c_\alpha \right )^2  \right ). 
\end{equation}       
         
Recordemos que la ecuación (B8) se dedujo a partir de la ecuaciones (B2) y (B3) y expresa la variación del  potencial debido a que se inducen variaciones en las concentraciones, $\Delta c_\beta $ =-$\Delta c_\alpha $, gracias a la sustitución $ \alpha \rightarrow \beta $, para especies atómicas $\alpha$ y $\beta$ diferentes que conforman la \emph{HEA}. 
  
 Teniendo en cuenta esto a partir de la ecuación (B8) podemos conocer como es la variación del potencial elástico respecto a la variación de la concentración  $\Delta c_i$, esto es, $ \left.\frac{\Delta e}{\Delta c_i}  \right |_j  $,  donde la barra con el subíndice indica la sustitución binaria $ i \rightarrow j $ (ver figura 3), por lo tanto se tiene que,   

 \begin{equation*}
  \left.\frac{\Delta e \left (  c_i, c_j  \right )}{\Delta c_i}  \right |_j = q \lambda_{i h}^{2} \left (\Delta c_i -  2c_i \Delta c_i - \left (\Delta c_i \right )^2 \right ) \left.\frac{1}{\Delta c_i }  \right |_j +  q \lambda_{j h}^{2} \left (\Delta c_j -  2c_j\Delta c_j - \left (\Delta c_j \right )^2 \right ) \left.\frac{1}{\Delta c_i}  \right |_j
  \end{equation*}   
 \begin{equation*}       
 \hspace{0.5cm}+ q \sum_{\alpha \neq i,j,h}^{N} \lambda_{\alpha h}^{2} \left (\Delta c_\alpha -  2c_\alpha \Delta c_\alpha - \left (\Delta c_\alpha \right )^2  \right ) \left.\frac{1}{\Delta c_i}  \right |_j  - 2q  \lambda_{i h} \lambda_{j h} \left ( -c_i \Delta c_i + c_j \Delta c_i - \left (\Delta c_i \right )^2  \right ) \left.\frac{1}{\Delta c_i}  \right |_j 
 \end{equation*}
 \begin{equation}\tag{B9}
 \hspace{-4.6cm} - 2q \sum_{\scriptsize \begin{matrix} 
\alpha,\beta \neq i,j,h \\      
  \alpha \neq \beta   
\end{matrix}}^{N} \lambda_{\alpha h} \lambda_{\beta h} \left ( -c_\alpha\Delta c_\alpha + c_\beta \Delta c_\alpha - \left (\Delta c_\alpha \right )^2  \right ) \left.\frac{1}{\Delta c_i}  \right |_j. 
\end{equation}      
 
Ambas sumatorias en la ecuación (B9) desaparecen gracias a que ninguna varia respecto a $ \Delta c_i $, por lo tanto, 

 \begin{equation*} 
 \left.\frac{\Delta e \left (  c_i, c_j  \right )}{\Delta c_i}  \right |_j = q \lambda_{ih}^{2} \left (\Delta c_i -  2c_i\Delta c_i - \left (\Delta c_i \right )^2 \right ) \left.\frac{1}{\Delta c_i}  \right |_j    +  q \lambda_{jh}^{2} \left (\Delta c_j -  2c_j\Delta c_j - \left (\Delta c_j \right )^2 \right ) \left.\frac{1}{\Delta c_i}  \right |_j  
 \end{equation*} 
 
 \begin{equation}\tag{B10}
 \hspace{-2.5cm} - 2q  \lambda_{ih} \lambda_{jh} \left ( -c_i\Delta c_i + c_j \Delta c_i - \left (\Delta c_i \right )^2  \right ) \left.\frac{1}{\Delta c_i}  \right |_j.  
 \end{equation} 
  
 Reemplacemos $ \Delta c_j = -\Delta c_i  $, en el segundo término del lado derecho de la ecuación (B10), 
   
 \begin{equation*}     
 \left.\frac{\Delta e \left (  c_i, c_j  \right )}{\Delta c_i}  \right |_j = q \lambda_{ih}^{2} \left (\left.\frac{\Delta c_i}{\Delta c_i}  \right |_j -  2c_i \left.\frac{\Delta c_i}{\Delta c_i}  \right |_j - \left.\frac{\left (\Delta c_i \right )^2}{\Delta c_i}  \right |_j \right )
 \end{equation*}   
 \begin{equation*}  
  \hspace{3.2cm}  +q \lambda_{jh}^{2} \left (-\left.\frac{\Delta c_i}{\Delta c_i}  \right |_j +   2c_j \left.\frac{\Delta c_i}{\Delta c_i}  \right |_j - \left.\frac{\left (-\Delta c_i \right )^2}{\Delta c_i}  \right |_j \right )   
 \end{equation*}     
  
 \begin{equation}\tag{B11}
 \hspace{3.8cm} - 2q  \lambda_{ih} \lambda_{jh} \left ( -c_i \left.\frac{\Delta c_i}{\Delta c_i}  \right |_j + c_j \left.\frac{\Delta c_i}{\Delta c_i}  \right |_j - \left.\frac{\left (\Delta c_i \right )^2}{\Delta c_i}  \right |_j  \right ) .  
 \end{equation}     
    
Sí hacemos en la ecuación (B11) $ \left.\frac{\Delta c_i}{\Delta c_i}  \right |_j = \left.\frac{\Delta c_j}{\Delta c_j}  \right |_i=1 $, y $ \left.\frac{\left (\Delta c_i \right )^2}{\Delta c_i}  \right |_j \approx  \Delta c_i $, entonces se tiene que,         
      
 \begin{equation} \tag{B12}     
 \left.\frac{\Delta e \left (  c_i, c_j  \right )}{\Delta c_i}  \right |_j = q \lambda_{ih}^{2} \left (1 -  2c_i - \Delta c_i \right ) 
    +q \lambda_{jh}^{2} \left (-1 +   2c_j - \Delta c_i \right )   
  - 2q  \lambda_{ih} \lambda_{jh} \left ( -c_i  + c_j  - \Delta c_i  \right ).  
 \end{equation} 
 
 Reagrupando los términos de la ecuación (B12) y simplificando se tiene que,  
 \begin{equation}\tag{B13}      
 \left.\frac{\Delta e \left (  c_i, c_j  \right )}{\Delta c_i}  \right |_j = q \left ( \lambda_{ih}^{2} - \lambda_{jh}^{2} \right )+2q \left (c_j \lambda_{jh}^{2} - c_i \lambda_{ih}^{2} - \lambda_{ih} \lambda_{jh} \left ( c_j - c_i \right ) \right )- q\left ( \lambda_{ih}-  \lambda_{jh}  \right )^{2} \Delta c_i.  
 \end{equation}

Análogamente a partir de las ecuaciones (B2) y (B3), podemos conocer como es la variación del potencial respecto a la variación de la concentración $\Delta c_j $, con la sustitución binaria $j \rightarrow i $, resultando,  

 \begin{equation*}       
 \left.\frac{\Delta e \left (  c_i, c_j  \right )}{\Delta c_j}  \right |_i = -q \left ( \lambda_{ih}^{2} - \lambda_{jh}^{2} \right )-2q \left (c_j \lambda_{jh}^{2} - c_i \lambda_{ih}^{2} - \lambda_{ih} \lambda_{jh} \left ( c_j - c_i \right ) \right )  
 \end{equation*}
 \begin{equation}\tag{B14} 
   \hspace{-3cm} - q\left ( \lambda_{ih}-  \lambda_{jh}  \right )^{2} \Delta c_j,  
 \end{equation}
 con lo cual se deduce fácilmente que,
$\left.\frac{\Delta e \left (  c_i, c_j  \right )}{\Delta c_j}  \right |_i+\left.\frac{\Delta e \left (  c_i, c_j  \right )}{\Delta c_j}  \right |_i =0$. 

\section{Apéndice C}
A continuación se muestra el uso del software implementado con la ayuda de FORTRAN90, se escogieron un total de 60 elementos químicos en su mayoría metales de transición: $Li, Be, C, Na, Mg, Al, Si, K, Ca, Sc, Ti, V, Cr, Mn, Fe, Co, Ni, Cu, Zn, Ga, Ge, Sr, Y, Zr, Nb, Mo,$ $Ru, Rh, Pd, Ag, Cd, In, Sn, Sb, Ba, La, Ce, Pr, Nd, Sm, Eu, Gd, Tb, Dy, Ho, Er, Tm, Yb, Lu, Hf,$ $Ta, W, Re, Os, Ir, Pt, Au, Tl, Pb, Bi$. Hubieron algunas propiedades físico-químicas que no se obtuvieron en la bibliografía citada, las referencias se   encuentran dentro del mismo programa como comentarios. Se recomienda que el software sea compilado en Linux. En la carpeta donde se encuentra guardado el programa $HEA_{-} program.f90$, lo compilamos y lo ejecutamos vía terminal, por defecto esto genera el ejecutable $a.out$ el cual ejecutamos vía terminal,  apareciendo en pantalla las primeras imágenes que se muestran en las figuras C1 y C2,

\begin{figure}[!htb]   
\centering
\includegraphics[width=0.8\textwidth]{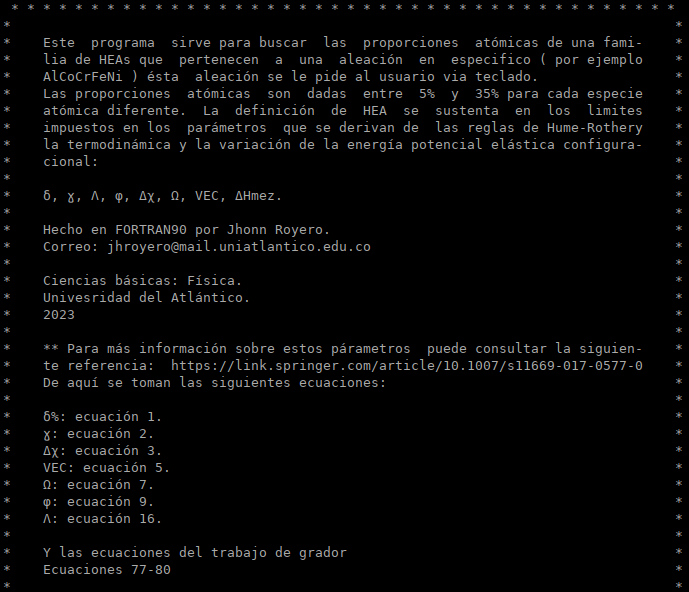} 
\begin{center} \small 
\emph{\textbf{Figura C1.} Primer texto que aparece en pantalla una vez ejecutado el archivo $a.out$. }  
\end{center} 
\end{figure}

\begin{figure}[!htb]   
\centering
\includegraphics[width=0.8\textwidth]{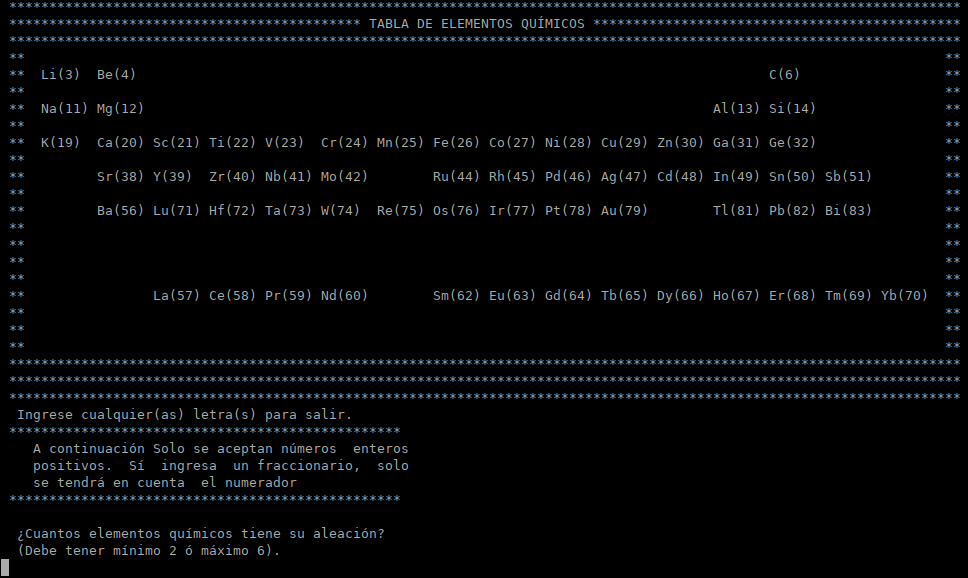} 
\begin{center} \small 
\emph{\textbf{Figura C2.} Texto final que aparece en pantalla una vez ejecutado el archivo $a.out$. }  
\end{center} 
\end{figure}

Una vez ejecutado el archivo $a.out$, se le pregunta al usuario: (i) ¿Cuantos elementos químicos se compone la aleación? (ii) ¿Respecto que elemento se va a graficar el eje X horizontal? por ejemplo el $Al$ es el elemento químico número 1 en la aleación $AlCoCrFeNi$ y el $Zr$ es el número 3 en la aleación $TiVZRNbHf$, (iii) el paso con el que desea trabajar en las proporciones molares, se puede escoger entre 0.01 y 0.001, ver figura (C3),

\begin{figure}[!htb]   
\centering
\includegraphics[width=0.7\textwidth]{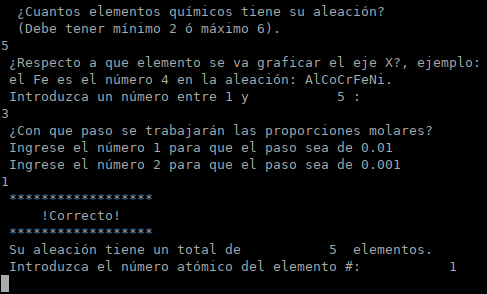} 
\begin{center} \small 
\emph{\textbf{Figura C3.} Condiciones que se piden de entrada, una vez ejecutado el archivo $a.out$. Al final de esta imagen se le pide al usuario que ingrese los números atómicos de los cuales se componen su aleación.   }   
\end{center} 
\end{figure}

\begin{figure}[!htb]   
\centering
\includegraphics[width=0.7\textwidth]{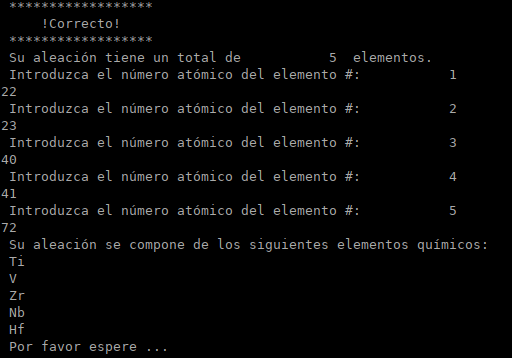} 
\begin{center} \small 
\emph{\textbf{Figura C4.} Condiciones que se piden de entrada, una vez ejecutado el archivo $a.out$. }  
\end{center} 
\end{figure}

Una vez se ingresan los números atómicos de los elementos químicos que conforman nuestra aleación (figura C4) el programa mostrará los símbolos químicos de dichos elementos, finalizando con un \emph{Por favor espere ...}, indicando que el programa está realizando los cálculos respectivos. Los documentos creados son los siguientes: \emph{rMin y rMax}, donde se guardo la información sobre los átomos de menor y de mayor radio atómico; \emph{Contador de HEAs}, como su nombre lo indica es un contador de aleaciones (o \emph{HEAs}); \emph{$\#$ de HEAs}, que nos muestra el número total de aleaciones que fueron analizadas; \emph{delta$\%$max}, que nos muestra un listado de menor a mayor valor de las posibles candidatas a $\Delta_{max}$; \emph{delta$_{-}$per}, es un archivo donde se guardó el valor de cada $\delta\%$ y la aleación a la que pertenece;  \emph{delta vs gamma}, nos muestra los parámetros $\delta$ vs $\gamma$; \emph{delta vs varEout2}, muestra el parámetro $\delta$ vs la ecuación (78); \emph{delta vs varEout3}, nos muestra a $\delta$ vs la ecuación (77);  \emph{electronegatividad}, muestra $\Delta\chi$ y la aleación a la que pertenece; \emph{|electronegatividad|max}, muestra un listado de las electronegatividades máximas; \emph{gamma} es un archivo que muestra el parámetro $\gamma$ y la aleación a la que pertenece; \emph{gammamax}, muestra un listado de los valores máximos de $\gamma$; \emph{HEAs}, es un archivo donde se enumeraron las aleaciones y sus respectivas proporciones molares; \emph{Hmez}, nos muestra la entalpía de formación dada por la ecuación (9) y la aleación a la que pertenece, \emph{|Hmez|max}, muestra un listado de los valores absolutos máximos de la entalpía de formación dada por la ecuación (9);  \emph{Lambda\%max}, muestra un lista de los valores máximos del parámetro  $\Lambda$ y la aleación a la que pertenece; \emph{Lambda$_{-}$per}, muestra un listado de los valores hallados para el parámetro $\Lambda$ y la aleación a la que pertenece; \emph{Omega}, muestra un listado de los valores hallados para el parámetro $\Omega$ y la aleación a la que pertenece, \emph{Omegamax},  muestra un lista de los valores máximos del parámetro  $\Omega$ y la aleación a la que pertenece;  \emph{Phi$_{-}$per}, muestra un listado de los valores hallados para el parámetro $\Phi$ y la aleación a la que pertenece;  \emph{Semaforo}, nos muestra si en cada paso se ha hallado información;  \emph{VEC} muestra la ecuación (7) y la aleación a la que pertenece; \emph{VECmax} muestra un lista de los valores máximos del parámetro  \emph{VEC} y la aleación a la que pertenece; \emph{X vs delta$_{-}$per}, muestra la concentración molar de la especie atómica anfitriona y el parámetro $\delta\%$, \emph{X vs electronegatividad}, muestra la concentración molar de la especie atómica anfitriona y el parámetro $\Delta\chi$; \emph{X vs gamma}, muestra la concentración molar de la especie atómica anfitriona y el parámetro $\gamma$; \emph{X vs Hmez}, muestra la concentración molar de la especie atómica anfitriona y la entalpía de formación $\Delta H_{mez}$ (ecuación 9); \emph{X vs H*mez},  muestra la concentración molar de la especie atómica anfitriona y la entalpía de formación $\Delta H_{mez}$ (ecuación 79); \emph{X vs H*mez=Ee-Ec}, muestra la concentración molar de la especie atómica anfitriona y la entalpía de formación $\Delta H_{mez}$ (ecuación 80); \emph{X vs Lambda$_{-}$per}, muestra la concentración molar de la especie atómica anfitriona y la entalpía de formación $\Lambda$; \emph{X vs Omega}, muestra la concentración molar de la especie atómica anfitriona y el parámetro $\Omega$ evaluado en la ecuación (9); \emph{X vs Omega(sVe)}, muestra la concentración molar de la especie atómica anfitriona y el parámetro $\Omega$ evaluado en la ecuación (79); \emph{X vs Omega(sVe+Ec)}, muestra la concentración molar de la especie atómica anfitriona y el parámetro $\Omega$ evaluado en la ecuación (80); \emph{X vs Phi}, muestra la concentración molar de la especie atómica anfitriona y el parámetro $\Phi$; \emph{X vs Scorr:Sid}, muestra la concentración molar de la especie atómica anfitriona y la relación $S_{corr}/S_{id}$ (ecuación 43); \emph{X vs Smez}, muestra la concentración molar de la especie atómica anfitriona y la entalpía de formación (ecuación 8); los archivos \emph{X vs sVe1, X vs sVe2, X vs sVe3, X vs sVe4, X vs sVe5, X vs sVe6}, muestran las concentraciones molares de la especie atómica anfitriona y la energía potencial elástica de cada especie atómica (ver figura 30 para la aleación $Al_xCoCrFeNi$); \emph{X vs VEC} muestra la concentración molar de la especie atómica anfitriona y el $VEC$ dado por la ecuación (7). De esta manera fueron creados los archivos para su respectivo estudio. El software se programó para que pudiera entregar un máximo de $2\cdot10^6$ aleaciones. Para las familia $AlCoCrFeNi$
 y $TiVzrNbHf$ con un paso de 0.01 en las concentraciones molares, se obtuvieron un total de 238030 aleaciones para cada familia.
 \end{document}